\documentstyle[preprint,aps,epsf,floats]{revtex}

\def\Slash#1{{#1\!\!\!\slash}}

\def\Dslash{D\!\!\!\!\slash}
\def\ppslash{p^{\,\prime}\!\!\!\!\!\slash}
\def\nslash{n\!\!\!\slash}
\def\bnslash{\bar n\!\!\!\slash}
\def\pslash{p\!\!\!\slash}
\def\qslash{q\!\!\!\slash}
\def\lslash{l\!\!\!\slash}

\def\Aslash{A\!\!\!\slash}

\def\OMIT#1{}

\newcommand{\nn}{\nonumber} 

\newcommand{\bn}{\bar n}
\newcommand{\bea}{\begin{eqnarray}}
\newcommand{\eea}{\end{eqnarray}}

\newcommand{\bnP}{\bn\!\cdot\! \hat P}

\begin{document}
\setlength\baselineskip{18pt}

\preprint{\tighten \vbox{\hbox{UCSD/PTH 00-28}
\hbox{hep-ph/0011336}
}}

\title{An effective field theory for collinear and soft gluons: \\[-5pt]
heavy to light decays}

\author{Christian W. Bauer$^{a}$, Sean Fleming$^b$,  
Dan Pirjol$^a$, and Iain W. Stewart$^a$\\[20pt]}

\address{\tighten 
${}^a$ Physics Department, University of California at San
Diego, La Jolla, CA 92093 \\[3pt]
${}^b$ Physics Department, Carnegie Mellon University, Pittsburgh, PA 15213\\
}

\maketitle

{\tighten
\begin{abstract} 

We construct the Lagrangian for an effective theory of highly energetic quarks
with energy $Q$, interacting with collinear and soft gluons.  This theory has
two low energy scales, the transverse momentum of the collinear particles,
$p_\perp$, and the scale $p_\perp^2/Q$.  The heavy to light currents are matched
onto operators in the effective theory at one-loop and the renormalization group
equations for the corresponding Wilson coefficients are solved.  This running is
used to sum Sudakov logarithms in inclusive $B \to X_s \gamma$ and $B \to X_u
\ell\bar\nu$ decays. We also show that the interactions with collinear gluons
preserve the relations for the soft part of the form factors for heavy to light
decays found by Charles et al., establishing these relations in the large energy
limit of QCD.

\end{abstract}

}
\vspace{0.7in}

\newpage

\section{Introduction}

The phenomenology of hadrons containing a single heavy quark is
greatly simplified by the fact that nonperturbative hadronic physics
can be parameterized by an expansion in $\Lambda_{\rm QCD}/m$, where
$m$ is the mass of the heavy quark.  At lowest order, interactions are
insensitive to the heavy quark mass and spin, leading to new
spin-flavor symmetries~\cite{IW}. These symmetries relate form factors
for decays of one heavy hadron to another heavy hadron. In
Ref.~\cite{IW,HQET} Heavy Quark Effective Theory (HQET) was
constructed as a general framework in which to explore heavy quark
physics.  The effective theory allows a systematic treatment of $1/m$
corrections and makes the symmetries manifest.  Inclusive decays of
heavy hadrons involving large momentum transfer to the decay products
can also be treated in HQET with the help of the operator product
expansion (OPE)\cite{incl}. At leading order the parton model results
are recovered, and nonperturbative corrections are parameterized by
matrix elements of higher dimensional operators, suppressed by powers
of $1/m$.

Decays of heavy hadrons to light hadrons cannot be treated exclusively with HQET
unless the four-momentum of the light degrees of freedom are small compared to
$m$. However, in regions of phase space where the light hadronic decay products
have large energy $E \sim m$, a different expansion in powers of $1/E$ can be
performed. In Ref.~\cite{dugan} Dugan and Grinstein used this approach to
construct the large energy effective theory (LEET), which describes the
interaction of very energetic quarks with soft gluons. However, LEET is missing
an important degree of freedom, namely collinear gluons, and does not reproduce
the IR physics of QCD~\cite{LEETincons}. In Ref.~\cite{BFL} it was shown that an
effective theory including both collinear and soft gluons correctly reproduces
the infrared physics of QCD at one loop.  This collinear-soft theory is needed
between the scale $E$ and an intermediate scale, below which collinear modes can
be integrated out.  For inclusive decays it was shown that the collinear-soft
theory can be matched at the intermediate scale onto a theory containing only
soft degrees of freedom.

The power counting in the collinear-soft theory is complicated by the presence
of two low energy scales, which must be properly accounted for. These scales can
be clearly seen by considering the momentum of a collinear quark. If the quark
moves along the light cone direction $n^\mu$ with momentum $Q\sim E\sim m$ then
$p=(p^+,p^-,{p}_\perp) \sim Q(\lambda^2,1,\lambda)$, where $\lambda$ is a small
parameter. Thus ${p}_\perp \sim Q\lambda$ is the intermediate scale.  With two
low energy scales it is more appropriate to count powers of $\lambda$ rather
than powers of $1/Q$~\cite{BFL}.  This is analogous to Non-Relativistic
QCD (NRQCD) for bound states of two heavy quarks, where one counts powers of the
velocity rather than powers of $1/m$~\cite{LMR}. Constructing such an effective
field theory bears some similarity to isolating momentum regimes using the
method of regions~\cite{BS} on full theory Feynman diagrams. There are, however,
advantages to using an effective field theory approach over the method of
regions; namely it is straightforward to systematically include power
corrections, and it is possible to properly account for operator running, which
sums Sudakov logarithms.  In order to consistently go beyond leading order it is
important to give a detailed construction of the effective field theory. This
was not done in Ref.~\cite{BFL}, and it is one of the main points of this paper.

The collinear-soft effective theory can be used to describe both inclusive and
exclusive heavy-to-light decays. For inclusive decays this theory is valid in
the regime where the phase space of the decay is restricted such that the final
hadronic state is forced to have low invariant mass and large energy.  This is
the case for large electron energy or small hadronic invariant mass in
semileptonic $B \to X_u \ell \overline{\nu}$ decays, and for large photon energy
in $B \to X_s \gamma$ decays. The Sudakov logarithms that appear in the endpoint
regions of these decays can be summed into the coefficient function of operators
by running in the collinear-soft theory between the scale $Q$ and $Q\lambda$,
and then running a soft operator from $Q\lambda$ to $Q\lambda^2$.  In
Ref.~\cite{BFL} Sudakov logarithms at the endpoint of the photon energy spectrum
in the decay $B \to X_s \gamma$ were summed in this manner. Here we sum Sudakov
logarithms between $Q$ and $Q\lambda$ for both $B \to X_s \gamma$ and $B\to X_u
\ell \overline{\nu}$. In the ratio of large moments of these decay rates effects
of physics below the intermediate scale cancel, and we reproduce previous
calculations carried out using the factorization formalism~\cite{Ira,LLR}.

It is also possible to apply the collinear-soft effective theory to exclusive
heavy-to-light decays. The form factors for such transitions have contributions
from the exchange of soft gluons with spectators (soft contributions), as well
as from the exchange of collinear gluons (so called hard contributions). In this
paper we will only consider the soft form factors, even though the two types of
contributions are believed to be the same order in
$1/m_b$~\cite{russian,french,beneke}. In Ref.~\cite{french} relations among the
soft form factors were derived, and it was shown that only three independent
functions are needed to describe heavy-to-light decays. However, these relations
were obtained within the framework of LEET, which does not include collinear
gluons. In this paper we show that the inclusion of collinear modes does not
alter the soft form factor relations to leading order in $\lambda$. Since the
collinear-soft effective theory reproduces the infrared physics of QCD at large
energies, this establishes these soft form factor relations in the large energy
limit of QCD.

In this paper we give a detailed construction of the collinear-soft effective
theory and apply it to general heavy to light decays. In section II the
Lagrangian for collinear gluons and collinear quarks is constructed and the
Feynman rules are given. The power counting for collinear gluons is formulated
in a gauge invariant way. The collinear-soft effective theory does not have the
same spin symmetry as LEET, but is still invariant under a helicity
transformation.  In section III we construct the heavy to light currents in the
effective theory at lowest order in $\lambda$. At this order the effective
theory current couples to an arbitrary number of collinear gluons with a
universal Wilson coefficient. The one-loop matching for the Wilson coefficients
are then derived. In section IV the renormalization group evolution of these
coefficients are computed. Finally, in section V we present two applications of
this effective theory. First we sum Sudakov logarithms in the ratio of large
moments of $B \to X_s \gamma$ and $B \to X_u \ell \bar\nu$ decay rates. Next we
show that in the collinear-soft theory only three independent soft form factors
describe exclusive heavy to light decays, establishing these form factor
relations in the large energy limit of QCD.  The one loop matching onto currents
in the effective theory allows us to calculate the perturbative corrections to
these form factor relations in an infrared safe manner. For the ratio of full
theory form factors these hard corrections agree with Ref.~\cite{beneke}.

\section{The effective theory} \label{eft}

Decays of heavy hadrons to highly energetic light hadrons are most conveniently
studied in the rest frame of the heavy hadron. In this reference frame the light
particles move close to the light cone direction $n^\mu$ and their dynamics is
best described in terms of light cone coordinates $p = (p^+, p^-, p_\perp)$,
where $p^+ = n \cdot p$, $p^- = \bn \cdot p$. For motion in the $z$ direction we
take $n^\mu=(1,0,0,-1)$ and $\bn^\mu=(1,0,0,1)$, so $\bn\cdot n=2$.  For large
energies the different light cone components are widely separated, with $p^-
\sim Q$ being large, while $p_\perp$ and $p^+$ are small. Taking the small
parameter to be $\lambda \sim p_\perp/p^-$ we have
\begin{equation} \label{scaling}
 p^\mu = \bn \cdot p\, \frac{n^\mu}{2} + (p_\perp)^\mu + 
   n\cdot p\,\frac{\bn^\mu}{2} = {\cal O}(\lambda^0) +{\cal O}(\lambda^1) +
   {\cal O}(\lambda^2) \,,
\end{equation}
where we have used $p^+p^- \sim p_\perp^2 \sim \lambda^2$ for fluctuations near
the mass shell.  The collinear quark can emit either a soft gluon or a gluon
collinear to the large momentum direction and still stay near its mass
shell. Collinear and soft gluons have light cone momenta that scale like
$k_c = Q(\lambda^2, 1,\lambda)$ and $k_s = Q(\lambda^2,\lambda^2,\lambda^2)$,
respectively. For scales above the typical off-shellness of the collinear
degrees of freedom, $k_c^2 \sim (Q \lambda)^2$, both gluon modes
are required to correctly reproduce all the infrared physics of the full
theory. This was described in \cite{BFL}, where it was shown that at a scale
$\mu\sim Q$ QCD can be matched onto an effective theory that contains heavy
quarks and light collinear quarks, as well as soft and collinear gluons.

The Lagrangian describing the interaction of collinear quarks with
soft and collinear gluons can be obtained at tree level by expanding
the full theory Lagrangian in powers of $\lambda$.  We start from the
QCD Lagrangian for massless quarks and gluons
\begin{eqnarray} \label{QCD}
 {\cal L}_{\rm QCD} &=& \bar \psi\: i \Dslash\: \psi 
  -\frac14 G_{\mu\nu} G^{\mu\nu} \,,
\end{eqnarray}
where the covariant derivative is $D_\mu = \partial_\mu - i g T^a A^a_\mu$, and
$G_{\mu\nu}$ is the gluon field strength.  We begin by removing the large
momenta from the effective theory fields, similar to the construction of
HQET~\cite{HQET}. In HQET there are two relevant momentum scales, the mass of
the heavy quark $m$ and $\Lambda_{\rm QCD}$. The scale $m$ is separated from
$\Lambda_{\rm QCD}$ by writing $p=m v+k$, where $v^2 = 1$ and the residual
momentum $k\ll m$.  The variable $v$ becomes a label on the effective
theory fields.  Our case is slightly more complicated because there are three
scales to consider. We split the momenta $p$ by taking
\begin{eqnarray}
  p = \tilde p + k \,, \qquad\mbox{where}\quad
  \tilde p\equiv \frac12 (\bn\cdot p)n + p_\perp \,.
\end{eqnarray}
The ``large'' parts of the quark momentum $\bn\cdot p\sim 1$ and $p_\perp\sim
\lambda$, denoted by $\tilde p$, become a label on the effective theory field,
while the residual momentum $k^\mu \sim \lambda^2$ is dynamical. This is
analogous to NRQCD where there are also three relevant scales $m$, $m\beta$, and
$m\beta^2$ (and $\beta\ll 1$ is the $q\bar q$ bound state velocity).  In NRQCD
the three scales can be separated \cite{LMR} by writing $P=(m,\vec 0) + p + k$
where $p\sim m\beta$ and the residual momentum $k\sim m\beta^2$.  In this case
both the momenta of order $m$ (i.e. $(1,\vec 0)$), and the momentum of order
$m\beta$ are labels on the effective theory fields.

The large momenta $\tilde p$ are removed by defining a new field
$\psi_{n,p}$ by
\begin{eqnarray}\label{xidef}
 \psi(x) = \sum_{\tilde p} e^{-i\tilde p\cdot x} \psi_{n,p}\,.
\end{eqnarray}
A label $p$ is given to the $\psi_{n,p}$ field, with the understanding that only
the components $\bn\cdot p$ and $p_\perp$ are true labels. Derivatives
$\partial^\mu$ on the field $\psi_{n,p}(x)$ give order $\lambda^2$
contributions. For a particle moving along the $n^\mu$ direction, the four
component field $\psi_{n,p}$ has two large components $\xi_{n,p}$ and two small
components $\xi_{\bn,p}$. These components can be obtained from the field
$\psi_{n,p}$ using projection operators
\begin{eqnarray}
 \xi_{n,p} = \frac{\nslash \bnslash}{4}\: \psi_{n,p}\,,\qquad
 \xi_{\bn,p} = \frac{\bnslash \nslash}{4}\: \psi_{n,p}\,,
\end{eqnarray}
and satisfy the relations
\begin{eqnarray}
& &\frac{\nslash \bnslash}{4}\: \xi_{n,p} = \xi_{n,p}\,, \quad \nslash\, 
  \xi_{n,p}=0\,,  \nn \\
& &\frac{\bnslash \nslash}{4}\: \xi_{\bn,p} = \xi_{\bn,p}\,, \quad 
  \bnslash\, \xi_{\bn,p}=0 \,.
\end{eqnarray}
In terms of these fields the quark part of the Lagrangian in
Eq.~(\ref{QCD}) becomes
\begin{eqnarray}\label{L_split}
 {\cal L} &=& \sum_{\tilde p, \tilde p'}\bigg[ \bar\xi_{n,p'} \frac{\bnslash}{2} 
 \Big( in\cdot D \Big) \xi_{n,p} 
 + \bar\xi_{\bn,p'} \frac{\nslash}{2}\Big( \bn\cdot p +  i\bn\cdot D\Big) 
   \xi_{\bn,p} \nn\\[4pt]
&&\qquad +\bar\xi_{n,p'} \Big(\Slash{p}_\perp + i \Dslash_\perp \Big) 
 \xi_{\bn,p} 
 + \bar\xi_{\bn,p'} \Big(\Slash{p}_\perp + i \Dslash_\perp \Big) \xi_{n,p}
 \bigg] \,.
\end{eqnarray}
Since the derivatives on the fermionic fields yield momenta of order $k \sim
\lambda^2$ they are suppressed relative to the labels $\bn\cdot p$ and
$p_\perp$. Without the $\bn\cdot D$ and $D_\perp$ derivatives, $\xi_{\bn,p}$ is
not a dynamical field. Thus, we can eliminate $\xi_{\bn,p}$ at tree level by
using the equation of motion
\begin{eqnarray} \label{frdefn}
  (\bn\cdot p + \bn\cdot iD) \xi_{\bn,p} = (\pslash_\perp +
i\Dslash_\perp ) \frac{\bnslash}{2} \xi_{n,p} \,.
\end{eqnarray}
This is similar to the approach taken in QCD quantized on the light
cone \cite{lightconeQCD} and in QCD in the infinite momentum frame
\cite{infiniteQCD}, where two components of the fermion field are
constrained auxiliary fields and are thus removed from the
theory. Eqs.~(\ref{L_split}) and (\ref{frdefn}) result in a Lagrangian
involving only the two components $\xi_{n,p}$:\footnote{\tighten Note
that Eq.~(\ref{preLc}) still includes particle/antiparticle and the
two spin degrees of freedom. However, on the light cone the spinor for
a spin-up (down) particle is identical to that of the spin-up (down)
anti-particle. See for example, Ref.~\cite{infiniteQCD}.}
\begin{eqnarray}  \label{preLc}
 {\cal L} &=& \sum_{\tilde p, \tilde p'} e^{-i(\tilde p - \tilde p')\cdot x}
  \bar\xi_{n,p'} \left[ n\cdot iD +(\pslash_\perp + i\Dslash_{\perp} )
  \frac{1}{\bn\cdot p + \bar n \cdot iD} (\pslash_\perp +i \Dslash_{\perp} )
  \right] \frac{\bnslash}{2} \xi_{n,p} \,.
\end{eqnarray}
Here the summation extends over all distinct copies of the fields labelled by
$\tilde p, \tilde p'$.  Note that the gluon field in $D^\mu$ includes collinear
and soft parts, $A^\mu \to A^\mu_c + A^\mu_s$. The two types of gluons are
distinguished by the length scales over which they fluctuate.  Fluctuations of
the collinear gluon fields $A^\mu_c$ are characterized by the scale $q^2\sim
\lambda^2$, while fluctuations of the soft gluon field $A^\mu_s$ are
characterized by $k^2\sim \lambda^4$. Since the collinear gluon field has large
momentum components $\tilde q\equiv (\bn\cdot q,q_\perp)$, derivatives acting on
these fields can still give order $\lambda^{0,1}$ contributions. To make this
explicit we label the collinear gluon field by its large momentum components
$\tilde q$, and extract the phase factor containing $\tilde q$ by redefining the
field: $A_c(x) \to e^{-i\tilde q \cdot x} A_{n,q}(x)$.  Inserting this into
Eq.~(\ref{preLc}) one finds
\begin{eqnarray} \label{pre2Lc}
  {\cal L} &=& \sum_{\tilde p, \tilde p', \tilde q} e^{-i(\tilde p-\tilde
  p')\cdot x} \bar\xi_{n,p'} \Bigg[ n\cdot iD\, +g e^{-i\tilde q\cdot x} n\cdot
  A_{n,q} + \Big( \pslash_\perp + i\Dslash_{\perp} + ge^{-i\tilde q\cdot x}
  \Aslash_{n,q}^\perp\Big) \\ 
  & & \qquad \times \frac{1}{\bn\cdot p + \bn\cdot
  iD + ge^{-i\tilde q\cdot x} \bn\cdot A_{n,q}} \Big(\pslash_\perp +i
  \Dslash_{\perp}+ ge^{-i\tilde q\cdot x} \Aslash_{n,q}^\perp \Big) \Bigg]
  \frac{\bnslash}{2} \xi_{n,p} \,.\nonumber
\end{eqnarray} 
Here the covariant derivative is defined to only involve soft
gluons. 

\begin{table}[t!]
\begin{center}
\begin{tabular}{l|c|c|c|cccc}
 & heavy quark & collinear quark\ & soft gluon\ && collinear gluons &
  \\\hline Field & $h_v$ & $\xi_{n,p}$ & $A_s^\mu$ & $\bar n\cdot
  A_{n,q}$ & $n\cdot A_{n,q}$ & $A_{n,q}^\perp$ & \\ Scaling &
  $\lambda^3$ & $\lambda$ & $\lambda^2$ & $\lambda^0$ & $\lambda^2$ &
  $\lambda$ &
\end{tabular}
\end{center}
{\tighten \caption{Power counting for the effective theory fields.}
\label{table_pc} }
\end{table}

\begin{figure}[!t]
\begin{eqnarray}
  && \begin{picture}(20,10)(20,0)
     \put(33,15){$(\tilde p,k)$}
     \mbox{\epsfxsize=3.4truecm \hbox{\epsfbox{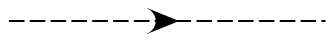}}  }
  \end{picture} \qquad\qquad\qquad\quad \Large 
  \raise5pt \hbox{$  = \ \mbox{\normalsize $i$}\, \frac{\nslash}{2}\: 
  {\bn\cdot p \over n\cdot k\, \bn\cdot p\: +\: p_\perp^2 +i\epsilon}$ } 
  \nn\\[-15pt]
  && \begin{picture}(20,80)(20,0)
     \mbox{\epsfxsize=3.4truecm \hbox{\epsfbox{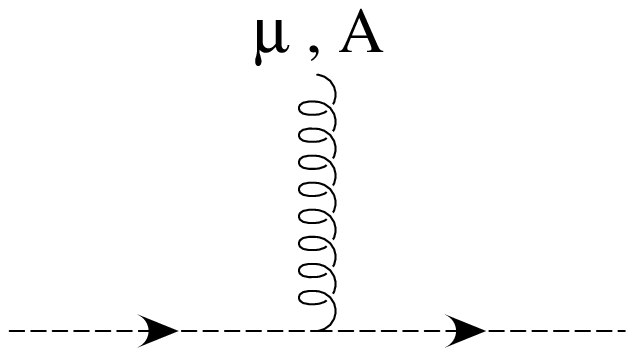}}  }
  \end{picture} \qquad\qquad\qquad\quad \Large 
  \raise20pt \hbox{$  = \mbox{\normalsize $i g\,T^A$} 
    \: n_\mu \, \frac{\bar\nslash}{2}  $ } \nn\\[10pt]
  && \begin{picture}(20,80)(20,0)
     \mbox{\epsfxsize=3.4truecm \hbox{\epsfbox{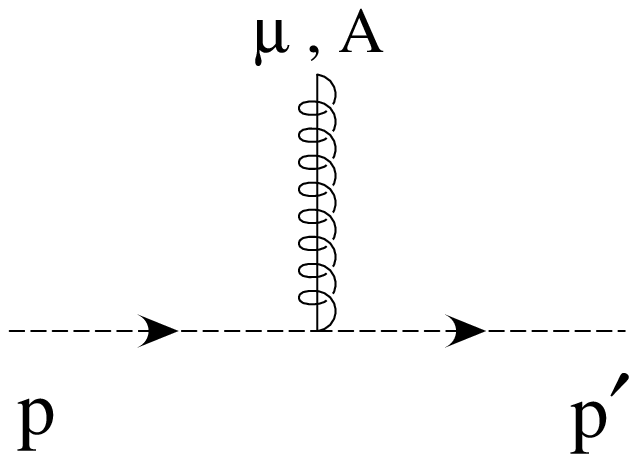}}  }
  \end{picture} \qquad\qquad\qquad\quad \Large 
  \raise50pt \hbox{$  = \mbox{\normalsize $i g\,T^A$} \, 
    \bigg[ n_\mu + \frac{ \gamma^\perp_\mu \pslash_\perp }{\bn \cdot p} 
    + \frac{ \ppslash_\perp \gamma^\perp_\mu }{\bn \cdot p^{\,\prime}}
    - \frac{\ppslash_\perp \pslash_\perp   }{\bn \cdot p\:  
    \bn \cdot p^{\,\prime}}\bar n_\mu \bigg]\, \frac{\bar\nslash}{2}  $ } 
 \nn\\[5pt]
  && \begin{picture}(20,80)(20,0)
     \mbox{\epsfxsize=3.4truecm \hbox{\epsfbox{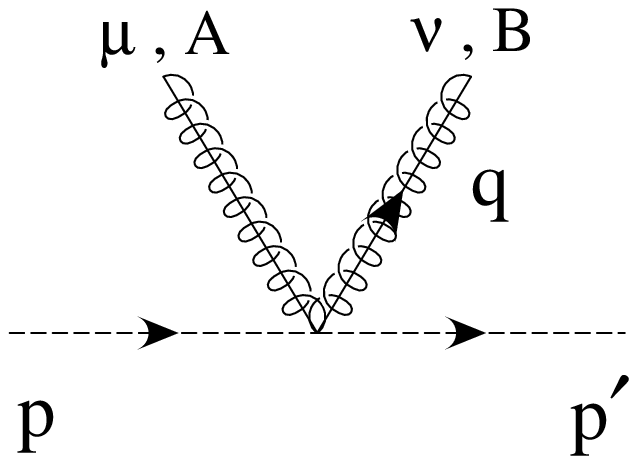}}  }
  \end{picture} \qquad\qquad\qquad\quad \large 
  \raise55pt \hbox{$  =\frac{i g^2\,T^A\,T^B}{ \bn \cdot (p-q)}\,  
    \bigg[ \gamma^\perp_\mu  \gamma^\perp_\nu  
    - \frac{ \gamma^\perp_\mu \pslash_\perp }{\bn \cdot p} \bn_\nu
    - \frac{ \ppslash_\perp \gamma^\perp_\nu }{\bn \cdot p^{\,\prime}}\bn_\mu
    + \frac{\ppslash_\perp \pslash_\perp  }{\bn \cdot p\:  
    \bn \cdot p^{\,\prime}} \bn_\mu \bn_\nu\bigg] \frac{\bar\nslash}{2}$ } 
 \nn\\[-40pt]
&& \hspace{1.43in} \large \raise55pt \hbox{$+\, \frac{i g^2\,T^B\,T^A}
    { \bn \cdot (q+p')}\, \bigg[ \gamma^\perp_\nu  \gamma^\perp_\mu  
    - \frac{ \gamma^\perp_\nu \pslash_\perp }{\bn \cdot p} \bn_\mu 
    - \frac{ \ppslash_\perp \gamma^\perp_\mu }{\bn \cdot p^{\,\prime}}\bn_\nu 
    + \frac{\ppslash_\perp \pslash_\perp }{\bn \cdot p\:  
    \bn \cdot p^{\,\prime}}\bn_\mu \bn_\nu \bigg]\frac{\bar\nslash}{2} $ } 
 \nn\\[-60pt] \nn
\end{eqnarray}
{\tighten \caption[1]{Order $\lambda^0$ Feynman rules: collinear quark
propagator with label $\tilde p$ and residual momentum $k$, and collinear quark
interactions with one soft gluon, one collinear gluon, and two collinear gluons
respectively.}
\label{fr1} }
\end{figure}

Finally, we expand Eq.~(\ref{pre2Lc}) in powers of $\lambda$.  To simplify the
power counting we follow the procedure~\cite{rescale} of moving all the
dependence on $\lambda$ into the interaction terms of the action to make the
kinetic terms of order $\lambda^0$. This is done by assigning a $\lambda$
scaling to the effective theory fields as given in Table~\ref{table_pc}. The
power counting in Table~\ref{table_pc} gives an order one kinetic term for
collinear gluons in an arbitrary gauge.  In generalized covariant gauge
\begin{eqnarray}
 \int d^4x\: e^{ik\cdot x}\: \langle 0 |\: T\: A_c^\mu(x) A_c^\nu(0)\: 
 | 0\rangle =\frac{-i}{k^2} \Big( g^{\mu\nu} - \alpha \frac{k^\mu
 k^\nu}{k^2} \Big)
\end{eqnarray}
and the scaling of the components on the right and left hand side of this
equation agree.\footnote{\tighten We have chosen a different counting for the
collinear gluon fields than Ref.~\cite{BFL} (where $A_c^\mu\sim \lambda$).  In
Feynman gauge there is the freedom to choose any scaling with $A_c^+ A_c^- \sim
\lambda^2$ (including the choice as in Ref.~\cite{BFL}). The choice in
Table~\ref{table_pc} is preferred since $A^\mu_c$ scales the same way as a
collinear momentum and there are no interactions that scale as $1/\lambda$.}
With this power counting all interactions scale as $\lambda^n$ with $n\ge 0$.
Expanding Eq.~(\ref{pre2Lc}) to order $\lambda^0$ gives
\begin{eqnarray}\label{Lc} 
 {\cal L}_{cs} &=& \bar\xi_{n,p} \bigg( n\cdot iD + \frac{p_\perp^2}{\bn\cdot p} 
 \bigg) \frac{\bnslash}{2}\xi_{n,p} \\
 &&+
 \bar\xi_{n,p+q} \bigg[ gn\cdot A_{n,q} + g\Aslash_{n,q}^\perp
 \frac{\pslash_{\perp}}{\bn\cdot p} 
 + \frac{\pslash_{\perp}+\qslash_{\perp}}{\bn\cdot (p+q)} g\Aslash_{n,q}^\perp 
 -\frac{\pslash_{\perp}+\qslash_{\perp}}{\bn\cdot (p+q)} g\bn\cdot
 A_{n,q}\frac{\pslash_{\perp}}{\bn\cdot p} \bigg]
 \frac{\bnslash}{2}\xi_{n,p} \nn \\
 &&+ \ldots  + {\cal O}(\lambda) \,. \nn
\end{eqnarray} 
Summation over the labels $\tilde p,\tilde q$ is understood implicitly. The
ellipsis denote terms of the same order in the power counting with two or more
collinear gluon fields, and arise because we expanded Eq.~(\ref{pre2Lc}) in
powers of $g A_c$ to obtain the above expression. This expansion was necessary
to move the collinear gluon phase factor appearing in the denominator of
Eq.~(\ref{pre2Lc}) into the numerator. This allowed us to remove the large
momentum $\tilde q$ from the Lagrangian so that all covariant derivatives were
truly of ${\cal O}(\lambda^2)$. The method for including terms of higher order
in $\lambda$ should be obvious from our derivation. The first few Feynman rules
which follow from the $\lambda^0$ terms in ${\cal L}_{cs}$ are shown in
Fig.~\ref{fr1}.

The first term in Eq.~(\ref{Lc}) gives the propagator for the collinear quarks,
which does not change depending on whether it interacts with soft or collinear
gluons.  This is distinct from the situation in the method of regions~\cite{BS},
where one must determine the propagator on a case by case basis.  The
interaction with a soft gluon is obtained from the covariant derivative term in
Eq.~(\ref{Lc}). Also shown in Fig.~\ref{fr1} are the interactions with one and
two collinear gluons.  The collinear gluon interactions are label changing
unlike the interaction involving soft gluons. Since $\bn\cdot A_{n,q}\sim
\bn\cdot p$, Eq.~(\ref{Lc}) includes interactions of a collinear quark with an
arbitrary number of collinear gluons. In Fig.~\ref{fr1} only interactions
through ${\cal O} (g^2)$ are shown. Note that in the light-cone gauge $\bn\cdot
A_{n,q}=0$ these Feynman rules are the complete set, since interactions of a
collinear quark with three or more collinear gluons vanish. In this gauge
similar Feynman rules for collinear gluons have been obtained in the framework
of light cone QCD~\cite{lightconeQCD_FR}.  However, the Feynman rules in
Fig.~\ref{fr1} can be used in any gauge.

The LEET Lagrangian corresponds to the $\bar\xi_{n,p}\, \frac{\bnslash}{2}\,
n\cdot iD\, \xi_{n,p}$ term in Eq.~(\ref{Lc}) and is invariant under a SU(2)
symmetry~\cite{dugan,french} with generators $S^1=(\gamma^0\Sigma^1)/2$,
$S^2=(\gamma^0\Sigma^2)/2$, and $S^3=\Sigma^3/2$ where $\Sigma^i$ are the
standard rotation generators. The collinear soft Lagrangian ${\cal L}_{cs}$ has
less symmetry than LEET because terms with $\gamma_\perp^1 \gamma_\perp^2$
violate the transformations generated by $S^1$ and $S^2$.  However, ${\cal
L}_{cs}$ is still invariant under a $U(1)$, namely the helicity transformations
generated by $S^3$.  Since $S^3=\gamma^5 (1/2-\bnslash\nslash/4)$ and $\nslash
\xi_{n,p}=0$ the helicity transformation also corresponds to the chiral
transformation generated by $\gamma^5/2$.
 
To complete the construction of the effective theory we have to include heavy
quarks. This can be done by adding the usual HQET Lagrangian for the heavy quark
field $h_v$,
\begin{eqnarray}\label{HQET}
  {\cal L}_{\rm HQET} = \bar h_v \: i v\cdot D\: h_v \,.
\end{eqnarray}
The covariant derivative in Eq.~(\ref{HQET}) contains only the soft gluon field
because the heavy quark field does not couple to collinear gluons~\cite{BFL}.
This is discussed in more detail in the next section. 

\begin{figure}[!t]
 \centerline{\mbox{\epsfysize=2.0truecm \hbox{\epsfbox{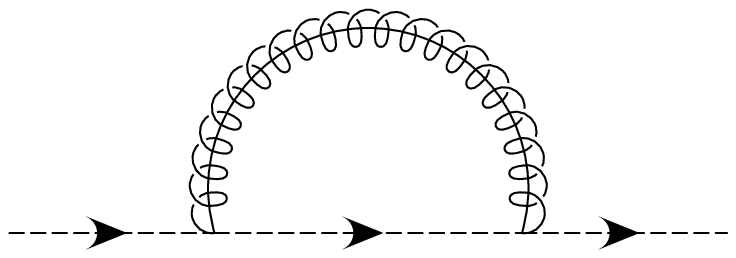}} \hspace{1.cm}
  \epsfysize=2.0truecm \hbox{\epsfbox{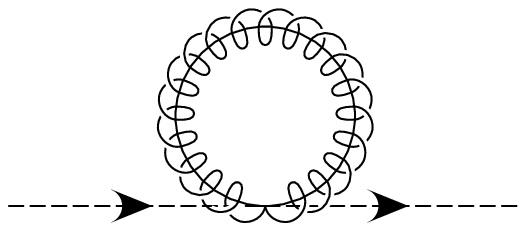}}  }}
 \medskip
{\tighten \caption[1]{Order $\alpha_s \lambda^0$ self energy diagrams for a 
collinear quark.} \label{fd1} }  
\end{figure}  
As a simple application of the Feynman rules consider the order $\lambda^0$
diagrams for the self energy of a collinear quark shown in Fig.~\ref{fd1}.  The
tadpole diagram vanishes in dimensional regularization.  In Feynman gauge the
remaining diagram gives
\begin{eqnarray} \label{self}
 i \Sigma_c(p) &=& g^2 C_F 
 \frac{\bar\nslash}{2}\,\int\frac{d^d l}{(2\pi)^d} \left\{
 (n\cdot\bar n)\frac{p_\perp^2+\pslash_\perp \lslash_\perp}{\bar n\cdot 
 p(p+l)^2 l^2}
 +(n\cdot\bar n)\frac{p_\perp^2+\lslash_\perp \pslash_\perp}{\bar n\cdot 
 p(p+l)^2 l^2} \right.\\
&+& \left. 2(d-4) \frac{p_\perp^2+l_\perp\cdot p_\perp}{\bar n\cdot p (p+l)^2 
 l^2} - (d-2)\left( \frac{(p_\perp+l_\perp)^2}{[\bar n\cdot (p+l)]^2} + 
 \frac{p_\perp^2}{[\bar n\cdot p]^2}\right) \frac{\bar n\cdot (p+l)}
 {(p+l)^2 l^2} \right\} \nn \,.
\end{eqnarray}
Here sums over the labels $\bn\cdot l$ and $l_\perp$ were combined with the
integrals over residual momenta to give the full $d^dl$ measure (c.f.,
Ref.~\cite{LMR}).  The first two terms in Eq.~(\ref{self}) correspond to the
$(\mu,\nu)=(+,-)$ and $(-,+)$ polarizations of the exchanged gluon, and the last
line to the $(\perp,\perp)$ contribution, respectively.  Computing the loop
integrals one finds
\begin{eqnarray}
 && i\Sigma_{+-}(p) + i\Sigma_{-+}(p) = \frac{i\alpha_s C_F}{4\pi}
 \frac{\bar\nslash}{2}\, \Gamma(\epsilon)
 \frac{\Gamma^2(1-\epsilon)}{\Gamma(2-2\epsilon)} \frac{2p_\perp^2}{\bar n\cdot
 p}\left(\frac{-p^2}{e^{\gamma_E} \mu^2} \right)^{-\epsilon} \\ 
&&
 i\Sigma_{\perp\perp}(p)= -\frac{i\alpha_s C_F}{4\pi} \frac{\bar\nslash}{2}\,
 \Gamma(\epsilon)\frac{\Gamma^2(1-\epsilon)}{\Gamma(2-2\epsilon)} \left\{
 (1+\epsilon)\,\frac{p_\perp^2}{\bar n\cdot p} - (1-\epsilon) n\cdot p\right\}
 \left(\frac{-p^2}{e^{\gamma_E} \mu^2}\right)^{-\epsilon}\,. \nn
\end{eqnarray}
Here (and in the rest of the paper) we use $\overline{\rm MS}$ and therefore
redefined $\mu^2 \to  \mu^2 e^{\gamma_E}/(4\pi)$.
The sum has precisely the form of the inverse collinear quark propagator in
Fig.~\ref{fr1}:
\begin{eqnarray}
 \Sigma_c(p) = \frac{\alpha_s C_F}{4\pi} \frac{\bar\nslash}{2}\,
 (1-\epsilon)\Gamma(\epsilon)
 \frac{\Gamma^2(1-\epsilon)}{\Gamma(2-2\epsilon)}
 \frac{p^2}{\bar n\cdot p}
 \left(\frac{-p^2}{e^{\gamma_E} \mu^2}\right)^{-\epsilon}\,.
\end{eqnarray}
The ultraviolet divergence in this expression is removed by onshell wavefunction
renormalization of the effective theory field $\xi_{n,p}$, 
\begin{eqnarray}
 Z_\xi = 1 - \frac{\alpha_s C_F}{4\pi}
 \left[\frac{1}{\epsilon} -\log\Big(\frac{-p^2}{\mu^2}\Big) + 1\right]\,.
\end{eqnarray}
$Z_\xi$ coincides with the renormalization of the quark field in QCD. This is
expected~\cite{BFL} since without currents or soft effects the collinear quark
Lagrangian simply describes QCD in a particular frame. The utility of the two
component formalism in Eq.~(\ref{Lc}) will become evident in the next section
where heavy to light currents are discussed.

\section{Matching the heavy to light currents} \label{match}

At a scale $\mu\sim Q$ the weak Hamiltonian has heavy to light 
semileptonic or radiative operators of the form~\cite{GSWS}
\begin{eqnarray} \label{fullH}
  H_{eff} = \frac{G_F}{\sqrt2}\:  V\: C^{\rm full}(\mu)\: 
  J_{\rm had} \: J \,,
\end{eqnarray}
where $V$ is the CKM factor, $J$ is a non-hadronic current, and the Wilson
coefficients $C^{\rm full}(\mu)$ have been run from the scale $\mu=m_W$ down to
$m_b$.  In Eq.~(\ref{fullH}), the hadronic currents are of the form $J_{\rm
had}=\bar q\, \Gamma\, b$ and we will consider $\Gamma = \{1$, $\gamma_5$,
$\gamma_\mu$, $\gamma_\mu\gamma_5$, $\sigma_{\mu\nu}$, $\sigma_{\mu\nu}\gamma_5
\}$. We choose this over-complete basis to simplify the treatment of $b\to
s\gamma$.  Below the scale $Q\sim \bn\cdot p$ the hadronic current can be
matched onto currents in the collinear-soft effective theory.  This introduces a
new set of Wilson coefficients $C_i(\mu)$. In this section the one loop matching
for these new coefficients will be performed at $\mu=m_b$, while the running
will be considered in section~IV. We could equally well match at $\mu=\bn\cdot
p$, but the difference is irrelevant since we treat $\bn\cdot p\sim m_b$ and do
not attempt to sum logarithms of the form $\ln(\bn \cdot p/m_b)$.

Naively, one might expect that at lowest order the effective theory hadronic
current is $J^{\rm eft}_{\rm had}= C(\mu)\bar\xi_{n,p}\:\Gamma\, h_v$. However,
since the label $\bn\cdot p\sim \lambda^0$, the effective theory Wilson
coefficient can also be a function of $\bn\cdot p$. Furthermore, an arbitrary
number of fields $\bn\cdot A_{n,q}\sim \lambda^0$ can be included without
additional power suppression.  At lowest order in $\lambda$ the most general
heavy to light current in the effective theory therefore has the form
\begin{eqnarray} \label{J1}
 J^{\rm eft}_{\rm had} &=&  c_0(\bn\cdot p,\mu)\: \bar\xi_{n,p}\:\Gamma\, h_v 
 \:+ c_1(\bn\cdot p,\bn\cdot q_1,\mu)\: \bar\xi_{n,p}\:(g\, \bn\cdot A_{n,q_1}) 
 \Gamma\, h_v \nn\\
 && + c_2(\bn\cdot p,\bn\cdot q_1,\bn\cdot q_2,\mu)\: \bar\xi_{n,p}\:
 (g\, \bn\cdot A_{n,q_1})(g\, \bn\cdot A_{n,q_2}) \Gamma\, h_v +\ldots \
 \,,
\end{eqnarray}
where the ellipsis stand for terms of the same order with more powers of
$\bn\cdot A_{n,q}$. The coefficients $c_i$ may also depend on the choice of
$\Gamma$.
\begin{figure}[!t]
 \centerline{\mbox{\epsfysize=4.0truecm \hbox{\epsfbox{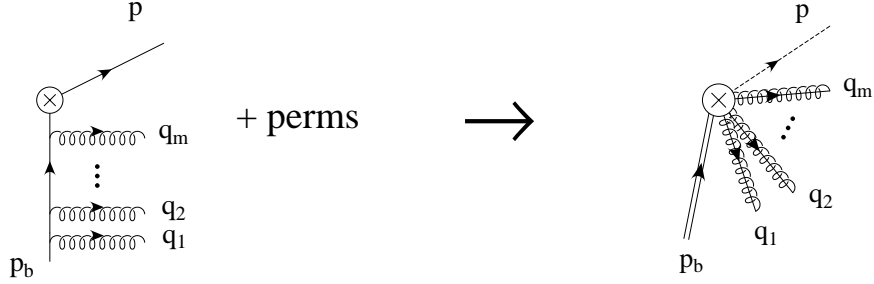}}  }}
{\tighten \caption[1]{Matching for the order $\lambda^0$ Feynman rule for the
heavy to light current with $n$ collinear gluons. All permutations of crossed 
gluon lines are included on the left.} 
 \label{fig_current} }  
\end{figure}
At the scale $\mu=m_b$ the $c_i$ can be determined by the tree level matching
calculation depicted in Fig.~\ref{fig_current}. On the left the gluons with
collinear momenta kick the $b$ quark far off-shell, and integrating out these
off-shell $b$ quarks gives the effective theory operator on the right.

To perform the matching, first consider the simpler case of an Abelian gauge
group. In this case calculating the full theory graph with $m$ gluons in
Fig.~\ref{fig_current}, expanding in powers of $\lambda$, and putting the result
over a common denominator gives
\begin{eqnarray} \label{cm}
  c_m(\mu=m_b) = \frac{1}{m!}\: \prod_{i=1}^m \frac{1}{\bn \cdot q_i} \,.
\end{eqnarray}
The factor of $1/m!$ is from the presence of $m$ identical $A_c$ fields at the
same point.\footnote{\tighten Note that in Ref.~\cite{BFL} the Feynman rule with
a single collinear gluon ($m=1$) has an additional $\nslash \Aslash_c /m_b$ term
which did not contribute to the results there. With the power counting in the
Table~\ref{table_pc} this term is suppressed by a power of $\lambda$.}  Thus,
we have the tree level result
\begin{eqnarray} \label{J2}
  J^{\rm eft}_{\rm had}\bigg|_{\mu=m_b} &=& \bar\xi_{n,p}\: 
 \exp\bigg( \frac{g\,\bn\cdot A_{n,q}}{\bn\cdot q} \bigg) 
 \:\Gamma\, h_v \,.
\end{eqnarray}
It is not immediately clear how this result is modified for $\mu< m_b$ since the
infinite series of operators in Eq.~(\ref{J1}) could each run
differently. However, gauge invariance relates these operators, and only the sum
of terms in Eq.~(\ref{J2}) is gauge invariant.  Under a collinear gauge
transformation $\alpha_{n,q}(x)$, the field $h_v$ is invariant since collinear
gluons do not couple to heavy quarks.  On the other hand, the collinear quark
field transforms as $\xi_{n,p} \to e^{i\alpha(x)} \xi_{n,p}$.  Thus, the
operator $\bar\xi_{n,p} \Gamma h_v$ is not gauge invariant. However, it is
straightforward to see that the operator in Eq.~(\ref{J2}) is invariant, and
this is done in Appendix~\ref{app_gauge}.  It is found that 
\begin{eqnarray}
 &&\exp\bigg(\frac{g\,\bn\cdot A_{n,q}}{\bn\cdot q} \bigg)  \to 
   \exp\bigg(\frac{g\,\bn\cdot A_{n,q}}{\bn\cdot q} \bigg)
   \exp\Big[ i\alpha(x) \Big] \,,
\end{eqnarray}
and the last exponential exactly cancels the transformation of $\bar\xi_{n,p}$.
By gauge invariance the current therefore has to be of the form in
Eq.~(\ref{J2}) for an arbitrary scale $\mu$.  It is convenient to define a field
that transforms as a singlet under a collinear gauge transformation
\begin{eqnarray}
  \chi_{n,P} = \exp\bigg(\frac{-g\,\bn\cdot A_{n,q}} {\bn\cdot q} \bigg) 
  \xi_{n,p}\,.
\end{eqnarray}
We will refer to $\chi_{n,P}$ as the jet field since it involves a collinear
quark field plus an arbitrary number of collinear gluons moving in the $n$
direction. The relevant label for the jet field is simply the sum of
labels of the particles in the jet, $P=p+\sum q_i$.  In terms of this field the
leading order effective theory current for $Q\lambda < \mu < m_b$ has the form
\begin{eqnarray}   \label{eftJ}
 J^{\rm eft}_{\rm had} &=& C_i(\mu,\bn\cdot P)\ \bar\chi_{n,P}\:
 \Gamma\, h_v \,,
\end{eqnarray}
with a universal coefficient $C_i(\mu,\bn\cdot P)$. The statement that the
coefficient only depends on the total jet momentum $P$ is
non-trivial and is discussed further in Appendix~\ref{app_scren}. 
\begin{figure}[!t]
\begin{eqnarray}
  && \begin{picture}(20,10)(20,0)
     \mbox{\epsfxsize=4.0truecm \hbox{\epsfbox{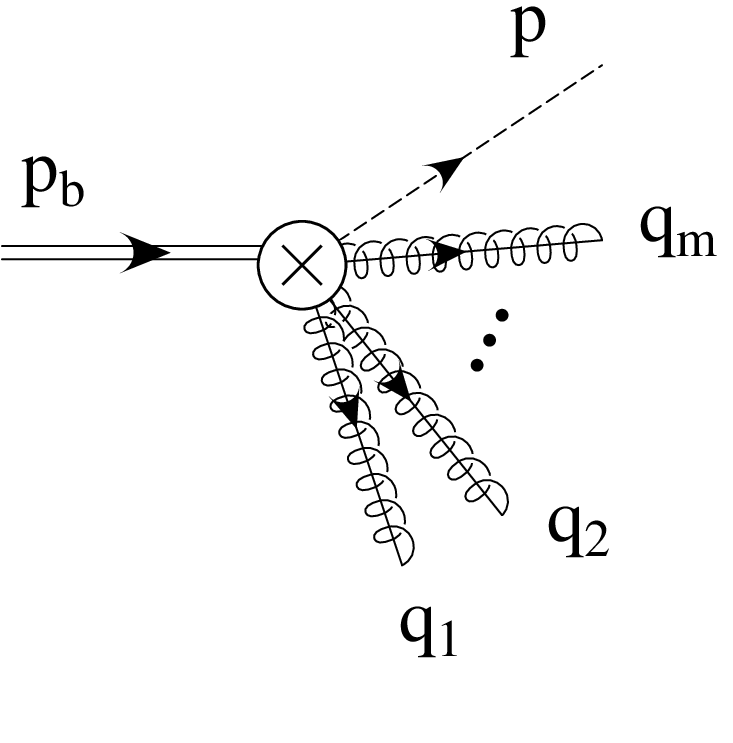}}  }
  \end{picture} \qquad\qquad\qquad\quad \Large 
  \raise60pt \hbox{$  =\mbox{\normalsize $i C(\mu,\bn\cdot P)\Gamma\:$}
  \mbox{\normalsize $g^m$} \begin{array}{c} \mbox{\Huge $\sum$} \\[-30pt] 
  \mbox{\small perms} \end{array} \frac{(\bn^{\mu_m} T^{A_m})\cdots 
  (\bn^{\mu_1} T^{A_1})} {[\bn\cdot q_1] [\bn\cdot (q_1+q_2)]\cdots
  [\bn\cdot \sum_{i=1}^m q_i]} $}  \nn
\end{eqnarray}
{\tighten \caption[1]{Order $\lambda^0$ Feynman rule for the effective theory
heavy to light current with $m$ collinear gluons. The sum is over permutations
of $\{1,\ldots,m\}$ and the Wilson coefficient depends only on the sum of momenta
in the jet, $P=p+\sum_{i=1}^m q_i$. }  \label{fig_fr2} }
\end{figure}

For a non-abelian gauge group a similar gauge invariance argument applies,
however the matching in Fig.~\ref{fig_current} is more
complicated. Eq.~(\ref{eftJ}) remains valid, but with a more complicated
definition of the jet field.  In momentum space we find
\begin{eqnarray} \label{jetq}
 \chi_{n,P} &=& \sum_k \sum_{\rm perms}
  \frac{(-g)^k}{k!}\left( {\bar n\cdot A_{\bn, q_1}\cdots
  \bar n\cdot A_{\bn, q_k} \over
  [\bn\cdot q_1] [\bn\cdot (q_1+q_2)]\cdots[\bn\cdot \sum_{i=1}^k q_i] }
  \right) \xi_{n,p} \,,
\end{eqnarray}
where the permutation sum is over the indices $(1,2,\dots,k)$. The
Feynman rules which follow from Eqs.~(\ref{eftJ}) and (\ref{jetq}) are shown in
Fig.~\ref{fig_fr2}. In position space the jet field takes the form of a
path-ordered exponential
\begin{eqnarray} \label{jetx}
 \chi_{n}(0) = \mbox{P}\exp\left(-ig\int_{-\infty}^0 \mbox{d}s\, 
 \bar n\cdot A^c(s \bar n^\mu ) \right)  \xi_n(0) \,,
\end{eqnarray}
where $P$ denotes path ordering along the light-like line collinear to $\bar
n$.\footnote{\tighten Path-ordered exponentials are also introduced to sum up
the couplings of soft gluons to a collinear jet, see Ref.~\cite{path}.}

In the effective theory both heavy and light quarks are described by two
component spinors, so there are only four heavy to light currents at leading
order in $\lambda$. We choose the linearly independent set $[\bar \chi_{n,P}\,
h_v]$, $[\bar \chi_{n,P}\, \gamma_5\, h_v]$, and $[\bar \chi_{n,P}\,
\gamma_\perp^\mu h_v]$, where $\gamma_\perp^\mu\! = \!\gamma^\mu\! -
\!n^\mu\bnslash/2\!-\!\bn^\mu\nslash/2$ has only two non-zero terms. The
matching of the heavy to light currents $\bar q\Gamma b$ onto operators in the
effective theory is
\begin{eqnarray} \label{ctrnsfm}
\bar qb &\to&  C_1(\mu)\, [\bar \chi_{n,P}\, h_v ] \,,  \\[3pt]
\bar q\gamma_5 b &\to&  C_2(\mu)\, [\bar \chi_{n,P}\, \gamma_5 h_v] \,,\nn\\[3pt]
\bar q\gamma_\mu b &\to& C_3(\mu)\, [\bar \chi_{n,P}\, \gamma_\mu^\perp h_v] +
   \Big\{ C_4(\mu)\, n_\mu +C_5(\mu)\, v_\mu\Big\} [\bar\chi_{n,P}\, h_v] 
   \nn\,, \\[4pt]
\bar q\gamma_\mu\gamma_5 b &\to& 
   C_6(\mu)\,i\epsilon_{\mu\nu}^\perp\,[\bar\chi_{n,P}\, \gamma^\nu_\perp h_v] -
   \Big\{ C_7(\mu)\,n_\mu +C_8(\mu)\,v_\mu\Big\} [\bar\chi_{n,P}\,\gamma_5 h_v]
  \,, \nn \\[4pt]
\bar q\, i\sigma_{\mu\nu} b &\to& 
  C_9(\mu)\,( n_\mu g_{\nu\lambda}\! -\!  n_\nu g_{\mu\lambda})
  [\bar\chi_{n,P}\, \gamma_\perp^\lambda h_v] 
  + C_{10}(\mu)\, i \epsilon^\perp_{\mu\nu} \, [\bar \chi_{n,P}\, \gamma_5 h_v]
  \nn \\*[0pt]
  &+& C_{11}(\mu)\, (v_\mu n_\nu -  v_\nu n_\mu)\, [\bar \chi_{n,P} h_v]
  + C_{12}(\mu)\,( v_\mu g_{\nu\lambda}\! -\! v_\nu g_{\mu\lambda}) 
  [\bar\chi_{n,P}\, \gamma_\perp^\lambda h_v] 
  \,,\nn\\[4pt]
\bar q\, i\sigma_{\mu\nu}\gamma_5 b &\to& - [C_{9}(\mu)+C_{12}(\mu)]\, 
  (i n_\mu\epsilon_{\nu\lambda}^\perp- i n_\nu\epsilon_{\mu\lambda}^\perp)  
  [\bar\chi_{n,P}\, \gamma_\perp^\lambda  h_v] 
  + C_{11}(\mu)\, i \epsilon^\perp_{\mu\nu} \, [\bar \chi_{n,P}\, h_v]
  \nn\\*[0pt]
 &+&  C_{10}(\mu)\,(v_\mu n_\nu-v_\nu n_\mu) [\bar \chi_{n,P} \gamma_5\, h_v]
 + C_{12}(\mu)\,(i v_\mu\epsilon_{\nu\lambda}^\perp- 
 i v_\nu\epsilon_{\mu\lambda}^\perp) [\bar \chi_{n,P} \gamma_\perp^\lambda h_v]
  \,, \nn 
\end{eqnarray}
where $\epsilon_{\mu\nu}^\perp= \epsilon_{\mu\nu\rho\sigma} v^\rho n^\sigma$
with $\epsilon_{0123}=-1$. From here on the dependence of the Wilson
coefficients on $\bn\cdot P$ will be suppressed. The relations in
Eq.~(\ref{ctrnsfm}) are valid\footnote{\tighten An exception is the relation
between the coefficients for $\bar q\, i\sigma_{\mu\nu} b$ and $\bar q\,
i\sigma_{\mu\nu}\gamma_5 b$ which can change depending on how $\gamma_5$ is
treated in $d$ dimensions.} to all orders in $\alpha_s$ and leading order in
$\lambda$.  At tree level the matching gives
\begin{eqnarray} \label{Ctree}
 C_{1,2,3,4,6,7,9,10,11}(m_b)=1\,,\qquad C_{5,8,12}(m_b)=0 \,.
\end{eqnarray}

To match these coefficients at one-loop, we calculate perturbative matrix
elements in the full and effective theories. All the long distance physics is
reproduced in the effective theory, and the difference between the two
calculations determines the short distance Wilson coefficients.  Since the
Wilson coefficients are universal the matching can be performed for the simpler
current $\bar\xi_{n,p} \Gamma h_v$ rather than the current $\bar\chi_{n,P}
\Gamma h_v$.  The calculation is most easily performed in pure dimensional
regularization.
\begin{figure}[!t]
 \centerline{ \mbox{
 \epsfysize=1.8truecm \hbox{\epsfbox{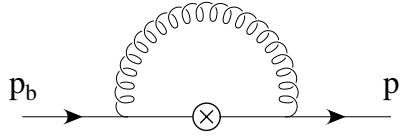}} }} \medskip 
{\tighten \caption[1]{Full theory one loop diagram for matching the heavy-light
current (denoted by $\otimes$). The incoming line is a
massive quark and the outgoing line is massless.} \label{fd2} }
\end{figure}  
The full theory matrix elements of the currents $\bar q\Gamma b$ between free
quark states are obtained by evaluating the diagram in Fig.~\ref{fd2} and
multiplying by the wavefunction and current renormalization factors. In
$d=4-2\epsilon$ dimensions the on-shell wavefunction renormalization constants
for massive and massless quarks are
\begin{eqnarray} \label{ZQ}
 Z_b = 1 + \frac{\alpha_s C_F}{4\pi}\left(
 -\frac{3}{\epsilon} + 3\ln\frac{m_b^2}{\mu^2} - 4\right)\,,\qquad
 Z_q = 1 \,,
\end{eqnarray}
and the renormalization constants for the scalar, pseudoscalar, vector,
axial vector, tensor and axial tensor currents are given by
\begin{eqnarray} \label{ZG}
 Z_S = Z_P = 1 - \frac{3\alpha_s C_F}{4\pi\epsilon}\,,\qquad
 Z_V = Z_A = 1\,,\qquad Z_T = Z_{T_5} = 1 + \frac{\alpha_s C_F}{4\pi\epsilon}\,.
\end{eqnarray}
The ultraviolet divergences in the $Z$'s cancel the ultraviolet divergences in
the diagram in Fig.~\ref{fd2}, hence all remaining $1/\epsilon$ divergences are
of infrared nature. The $b$ quark and light quark are taken to have momenta
$p_b$ and $p$ respectively, and we define $q=p_b-p$. Letting $\gamma_5$
anticommute in $d$ dimensions (the NDR scheme), the final result for the matrix
elements in the full theory is
\begin{eqnarray} \label{fullmatch}
\langle q| \bar{q} \{1,\gamma_5\} b | b \rangle &=& \Bigg\{ 1-\frac{\alpha_s
  C_F}{4 \pi} \Bigg[ \frac{1}{\epsilon^2} + \frac{5}{2 \epsilon} 
  + \frac{\ln\Big(\frac{\mu^2}{m_b^2}\Big)}{\epsilon} 
  - \frac{2\ln(1-\hat{q}^2)}{\epsilon} 
  + \frac{1}{2}\ln^2\Big(\frac{\mu^2}{m_b^2}\Big) \nn \\
&& -\frac{1}{2} \ln\Big(\frac{\mu^2}{m_b^2}\Big) 
  - 2 \ln(1-\hat{q}^2) \ln\Big(\frac{\mu^2}{m_b^2}\Big)
  + 2\ln^2(1-\hat{q}^2) \nn \\
&&  - \frac{2\ln(1-\hat{q}^2)}{\hat{q}^2} + 2 {\rm Li}_2(\hat{q}^2) 
  + \frac{\pi^2}{12}\: \Bigg] \Bigg\} \: \bar u \{1,\gamma_5\} u \,, \nn\\
\langle q| \bar{q}\{1,\gamma_5\} \gamma^\mu  b | b \rangle &=& \Bigg\{ 1
  -\frac{\alpha_s C_F}{4 \pi} \bigg[ \frac{1}{\epsilon^2} 
  + \frac{5}{2 \epsilon} + \frac{\ln\Big(\frac{\mu^2}{m_b^2}\Big)}
  {\epsilon} - \frac{2\ln(1-\hat{q}^2)}{\epsilon} + \frac{1}{2}
  \ln^2\Big(\frac{\mu^2}{m_b^2}\Big) \nn \\
&& + \frac{5}{2} \ln\Big(\frac{\mu^2}{m_b^2}\Big) - 2 \ln(1-\hat{q}^2)
  \ln\Big(\frac{\mu^2}{m_b^2}\Big) + 2\ln^2(1-\hat{q}^2) \nn \\ 
&& + \ln(1-\hat{q}^2) \Big(\frac{1}{\hat{q}^2}-3 \Big) + 
  2 {\rm Li}_2 (\hat{q}^2) + \frac{\pi^2}{12} + 6 \bigg] \Bigg\}\: 
 \bar u \{1,\gamma_5\}\gamma^\mu  u \nn \\
&& + \frac{\alpha_s C_F}{4 \pi} \bigg[ \frac{4}{\hat{q}^2} \ln(1-\hat{q}^2) 
  -\frac{2}{\hat{q}^2} -  \frac{2}{\hat{q}^4} \ln(1-\hat{q}^2) \bigg]\: 
  \hat{p}^\mu \, \bar u\{1,\gamma_5\} u \nn \\
&& + \frac{\alpha_s C_F}{4 \pi} \bigg[ \frac{2}{\hat{q}^2} 
   - \frac{2}{\hat{q}^2} \ln(1-\hat{q}^2) 
   + \frac{2}{\hat{q}^4}\ln(1-\hat{q}^2) \bigg]\: \hat{p}^\mu_b\, 
  \bar u \{1,\gamma_5\} u\,, \nn \\
\langle q| \bar{q} \{1,\gamma_5\}i \sigma^{\mu\nu}  b | b \rangle &=& \Bigg[
  1-\frac{\alpha_s C_F}{4 \pi} \Bigg[ \frac{1}{\epsilon^2} 
  + \frac{5}{2 \epsilon} + \frac{\ln\Big(\frac{\mu^2}{m_b^2}\Big)}
  {\epsilon} - \frac{2\ln(1-\hat{q}^2)}{\epsilon} 
  + \frac{1}{2}\ln^2\Big(\frac{\mu^2}{m_b^2}\Big) \nn \\
&&+ \frac{7}{2} \ln\Big(\frac{\mu^2}{m_b^2}\Big) - 2 \ln(1-\hat{q}^2)
   \ln\Big(\frac{\mu^2}{m_b^2}\Big) + 2\ln^2(1-\hat{q}^2) \nn \\ 
&& + 2 \ln(1-\hat{q}^2) \Big(\frac{1}{\hat{q}^2}-2 \Big) 
  +2 {\rm Li}_2 (\hat{q}^2) + \frac{\pi^2}{12} + 6 \Bigg] \Bigg\} \:
  \bar u\{1,\gamma_5\}\, i \sigma^{\mu\nu}  u \nn \\
&& + \frac{\alpha_s C_F}{4 \pi} \bigg[ \frac{4}{\hat{q}^2} \ln(1-\hat{q}^2)
  \bigg] \ \bar u  \{1,\gamma_5\} (\hat{p}^\mu \gamma^\nu  
 - \hat{p}^\nu \gamma^\mu) u \,, 
\end{eqnarray}
where the hat denotes momenta normalized with respect to $m_b$, so $\hat
q=q/m_b$.  This full theory result can be expanded in $\lambda$ by noting that
\begin{eqnarray} \label{q2}
 \hat q^2=1-\bn \cdot \hat p + {\cal O}(\lambda^2)\,,
\end{eqnarray}
and that at lowest order we can expand the full theory spinors using 
Eqs.~(\ref{ctrnsfm}) and (\ref{Ctree}).

For the effective theory in pure dimensional regularization the final collinear
quark is taken on-shell.  For momentum labels $(\bn\cdot p,p_\perp)$ this
corresponds to choosing this quarks residual momentum $k$ such that
$\bn\cdot p\ n\cdot k+p_\perp^2=0$.  In this case all graphs in the
effective theory are proportional to $1/\epsilon_{\rm UV}-1/\epsilon_{\rm
IR}=0$.  The ultraviolet divergences are canceled by effective theory
counterterms, and all infrared divergences cancel in the difference between the
full and effective theories. Thus, from Eq.~(\ref{fullmatch}) the Wilson
coefficients at the scale $\mu=m_b$ are
\begin{eqnarray} \label{matchC}
C_{1,2}(m_b) &=& 1 - \frac{\alpha_s(m_b)C_F}{4\pi} \Bigg\{ 2 \ln^2 (\bnP) 
  + 2 {\rm Li}_2(1\!-\!\bnP) - \frac{2\ln(\bnP)}{1-\bnP} +\frac{\pi^2}{12} 
  \Bigg\} \,, \nn \\*
C_{3,6}(m_b) &=& 1 - \frac{\alpha_s(m_b)C_F}{4\pi} \Bigg\{ 2\!\ln^2(\bnP) 
  + 2 {\rm Li}_2(1\!-\!\bnP) + \ln(\bnP) \left( \frac{3\bnP-2}{1-\bnP}\right)
  + \frac{\pi^2}{12} + 6\Bigg\} , \nn \\*
C_{4,7}(m_b) &=& 1 - \frac{\alpha_s(m_b)C_F}{4\pi} \Bigg\{ 2\!\ln^2(\bnP) 
  + 2 {\rm Li}_2(1\!-\!\bnP) - \ln(\bnP) \bigg[ \frac{2-4\bnP+(\bnP)^2}
  {(1-\bnP)^2}\bigg] \nn\\*
 && \quad +\frac{\bnP}{1-\bnP} + \frac{\pi^2}{12} + 6 \Bigg\} 
 \,, \nn \\*
C_{5,8}(m_b) &=& \frac{\alpha_s(m_b)C_F}{4\pi}\: \Bigg\{ \frac{2}{(1-\bnP)} 
  + \frac{2\bnP\ln(\bnP)} {(1-\bnP)^2} \Bigg\} \,,\\
C_{9}(m_b) &=& 1 - \frac{\alpha_s(m_b)C_F}{4\pi} \Bigg\{ 2\!\ln^2(\bnP) 
  + 2 {\rm Li}_2(1\!-\!\bnP) -2 \ln(\bnP) + \frac{\pi^2}{12} + 6 \Bigg\}
  \,,\nn\\
C_{10,11}(m_b) &=& 1 - \frac{\alpha_s(m_b)C_F}{4\pi} \left[ 2\!\ln^2(\bnP) 
  + 2 {\rm Li}_2(1\!-\!\bnP) +  \ln(\bnP) \left( \frac{4\bnP-2}{1-\bnP}
  \right) + \frac{\pi^2}{12} + 6 \right] , \nn \\
C_{12}(m_b) &=& 0  \,. \nn 
\end{eqnarray}
For the operator $\bar\xi_{n,p} \Gamma h_v$ there is only one particle in the
jet, so in that case $P=p$. In NDR the relations amongst Wilson coefficients, 
$C_1=C_2$, $C_3=C_6$, $C_4=C_7$, $C_5=C_8$, $C_{10}=C_{11}$, and $C_{12}=0$
hold true to all orders in perturbation theory for a massless light quark. This
is because the transformation, $q\to \gamma_5 q$ is a symmetry of massless QCD
and the U(1) helicity symmetry of Eq.~(\ref{Lc}) allows $\chi_{n,p}\to \gamma_5
\chi_{n,p}$, and these transformations relate currents with and without
$\gamma_5$.

\section{Renormalization group evolution} \label{run}

In this section we calculate the running of the Wilson coefficients in the
effective theory.  The coefficients mix into themselves and satisfy a
renormalization group equation of the form
\begin{eqnarray} \label{rge0}
  \mu \frac{d}{d\mu} C(\mu) = \gamma(\mu) C(\mu)\,.
\end{eqnarray}
The fact that Eq.~(\ref{rge0}) is homogeneous reproduces the exponentiation of
Sudakov logarithms. In this case it is natural to solve the renormalization
group equations for the quantity $\ln C(\mu)$.  The leading and subleading
series of logarithms are determined by the coefficients summarized in
Table.~\ref{table_logs}.
\begin{table}[t!]
\begin{center}
\begin{tabular}{clcc|cccc}
 &\multicolumn{2}{c}{series in $\ln C(\mu)$} && one-loop & 
  two-loops & three-loops \\\hline 
 & LL & $\alpha_s^n \ln^{n+1}$ && $1/\epsilon^2$ & -- & -- \\
 & NLL & $\alpha_s^n \ln^{n\phantom{+1}}$ && $1/\epsilon$ & $1/\epsilon^2$
  &--\\ 
 & NNLL &$\alpha_s^n \ln^{n-1}$ && matching & $1/\epsilon$ & 
  $1/\epsilon^2$
\end{tabular} 
\end{center}
{\tighten \caption{Coefficients in the effective theory loop graphs which we
anticipate are needed to predict the series of Sudakov logarithms in $\ln
C(\mu)$.}
\label{table_logs} }
\end{table}
From the Table one can see that for coefficients with tree level matching, the
one-loop matching in section~\ref{match} is not needed until NNLL order.

\begin{figure}[!t]
 \centerline{\mbox{\epsfysize=1.9truecm \hbox{\epsfbox{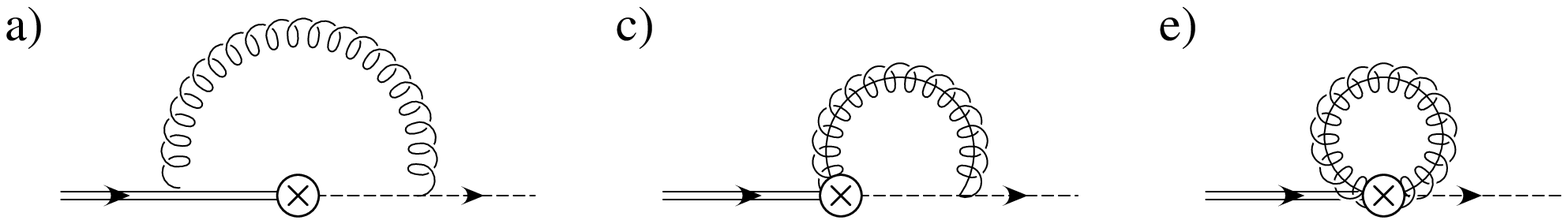}}} } \medskip
 \medskip\medskip\medskip  \centerline{\mbox{\epsfysize=1.5truecm
 \hbox{\epsfbox{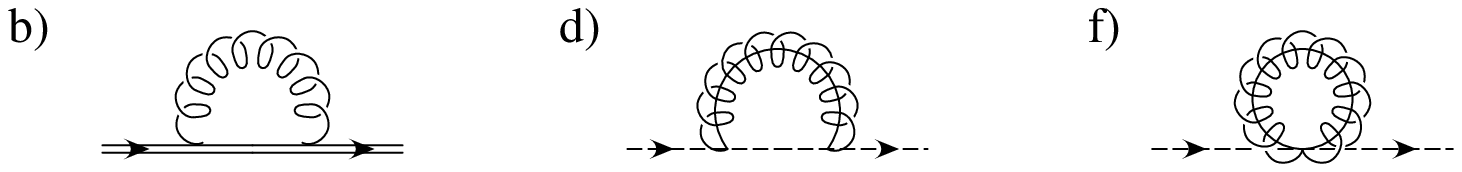} }} } \medskip 
{\tighten \caption[1]{Order $\lambda^0$ effective theory diagrams for the heavy
 to light current at one loop. }
\label{fd_eft} }  
\end{figure}

In section~\ref{match} it was shown that the coefficient of the effective theory
current $\bar\chi_{n,P} \Gamma h_v$ is the same as the current $\bar\xi_{n,p}
\Gamma h_v$, so only the renormalization of the simpler $\bar\xi_{n,p} \Gamma
h_v$ current needs to be considered. At one-loop the effective theory diagrams
are shown in Fig.~\ref{fd_eft}.  To distinguish UV and IR divergences we choose
the collinear quark momentum $p=\tilde p+k$ with label $\tilde p=(\bn\cdot
p,0,p_\perp)$ and zero residual momentum, $k=0$. In this case $p^2=p_\perp^2\ne
0$ and this off-shellness regulates IR divergences in the diagrams.  We will use
Feynman gauge.  The soft diagrams in Fig.~\ref{fd_eft} give
\begin{eqnarray} \label{efts}
 \mbox{Fig.\,\ref{fd_eft}a} &=& i\bar\xi_{n,p}\Gamma h_v  
 \frac{C_F\alpha_s(\mu)C(\mu)}{4\pi} \Bigg[ 
 -\frac{1}{\epsilon^2} 
 -\frac{2}{\epsilon}\ln\bigg(\frac{\mu\,\bn\cdot p}{-p_\perp^2\!-\!i\epsilon}
 \bigg) 
 - 2 \ln^2\bigg(\frac{\mu\,\bn\cdot p}{-p_\perp^2\!-\!i\epsilon}\bigg)
 - \frac{3\pi^2}{4} \Bigg] \,,\nn\\
 \mbox{Fig.\,\ref{fd_eft}b} &=& i v\cdot k \frac{\alpha_s(\mu)C_F}{4\pi}
 \Bigg[-\frac{2}{\epsilon}-4-4\ln\bigg(\frac{\mu}{-2v\cdot k\!-\!i\epsilon}
 \bigg)\Bigg]\,,
\end{eqnarray}
where $k$ is a residual momentum in the heavy quark wavefunction diagram.  The
order $\lambda^0$ soft wavefunction renormalization of the collinear quark is
not shown since in Feynman gauge it is proportional to $n^2=0$.  Evaluating the
diagrams with a collinear gluon in Fig.~\ref{fd_eft} gives
\begin{eqnarray} \label{eftc}
 \mbox{Fig.\,\ref{fd_eft}c} &=& i\bar\xi_{n,p}\Gamma h_v
  \frac{C_F\alpha_s(\mu)C(\mu)}{4\pi} \Bigg[ 
  \frac{2}{\epsilon^2}
  +\frac{2}{\epsilon} +\frac{2}{\epsilon}\ln\bigg(\frac{\mu^2}
  {-p_\perp^2\!-\!i\epsilon}\bigg) + \ln^2\bigg(\frac{\mu^2}
  {-p_\perp^2\!-\!i\epsilon} \bigg) \nn\\ && +
  2 \ln\bigg(\frac{\mu^2}{-p_\perp^2\!-\!i\epsilon}\bigg)
  +4-\frac{\pi^2}{6} \Bigg] \,,\nn\\ 
 \mbox{Fig.\,\ref{fd_eft}d} &=& \frac{i\bnslash}{2} \frac{p_\perp^2}{\bn\cdot p}
  \: \frac{\alpha_s(\mu)C_F}{4\pi} \Bigg[\frac{1}{\epsilon}+1 +
  \ln\bigg(\frac{\mu^2}{-p_\perp^2\!-\!i\epsilon} \bigg)\Bigg]\,, \nn\\ 
 \mbox{Fig.\,\ref{fd_eft}e,f} &=& 0 \,.
\end{eqnarray}
The graph in Fig.~\ref{fd_eft}d was calculated explicitly in section II.

From Eq.~(\ref{efts}) we see that the logarithms in diagrams with collinear
gluons are small at a scale $\mu\sim \sqrt{p_\perp^2} \sim Q\lambda$. For the
graphs with soft gluons the logarithms are small at a different scale $\mu\sim
p_\perp^2/(\bn\cdot p)\sim Q\lambda^2$.  Running the collinear-soft theory from
$\mu = Q$ to $\mu = Q\lambda$ therefore sums all logarithms originating from
collinear effects and part of the logarithms from soft exchange. At
$\mu=Q\lambda$ collinear gluons may be integrated out and one matches onto a
theory containing only soft degrees of freedom.  The running in this soft theory
includes the remaining logarithms from soft exchange, which would need to be
taken into account to sum all Sudakov logarithms.

To run between $Q$ and $Q\lambda$ we add up the ultraviolet divergences in the
soft and collinear diagrams in Eqs.~(\ref{efts}) and (\ref{eftc}).  This gives
the counterterm in the effective theory
\begin{equation}\label{Z_V}
 Z_i = 1 + \frac{\alpha_s(\mu) C_F}{4 \pi} \left[ \frac{1}{\epsilon^2} +
  \frac{2}{\epsilon}\ln\Big({\mu\over\bn\cdot P}\Big) 
  + \frac{5}{2 \epsilon} \right]\,.
\end{equation}
For $b\to s\gamma$, $\bn\cdot P=m_b$ and Eq.~(\ref{Z_V}) agrees with
Ref.~\cite{BFL}. Since $\mu > Q\lambda$ the counterterm can depend on the label
$\bn\cdot P\sim Q$, but does not depend on $P_\perp\sim Q\lambda$.  $Z_i$ could
also have been calculated directly from the matching result in
Eq.~(\ref{fullmatch}). Since the effective theory reproduces all the infrared
divergences in the full theory, the effective theory UV divergences are simply
the negative of the full theory IR divergences when pure dimensional
regularization is used.  This alternative approach also gives Eq.~(\ref{Z_V}).

In the effective theory the current $\bar\xi_{n,p} \Gamma h_v$ factors out of
the diagrams in Fig.~\ref{fd_eft} so it is obvious that $Z_i$ is
independent of the spin structure of the current.  Thus, all the coefficients
satisfy the same renormalization group equation (RGE)
\begin{eqnarray} \label{rge}
  \mu \frac{d}{d\mu} C_i(\mu) = \gamma(\mu) C_i(\mu)\,.
\end{eqnarray}
The LO anomalous dimension is determined by the $\ln(\mu)/\epsilon$ term in
Eq.~(\ref{Z_V}) (whose coefficient is determined by the $1/\epsilon^2$
term). The NLO anomalous dimension has a contribution from the $1/\epsilon$
terms in Eq.~(\ref{Z_V}), as well as a contribution from the $\ln(\mu)/\epsilon$
term in the two loop $Z_i$ counterterm:
\begin{eqnarray} \label{adim1}
 \gamma_{LO} &=& -\frac{\alpha_s(\mu) C_F}{\pi} 
  \ln\bigg( \frac{\mu}{\bn\cdot P} \bigg) \,, \nn\\
 \gamma_{NLO} &=& -\frac{5\alpha_s(\mu) C_F}{4 \pi} -2 C_F 
   B\, \frac{\alpha_s^2(\mu)}{(2\pi)^2}\,
   \ln\Big( \frac{\mu}{\bn\cdot P}\Big)\,. 
\end{eqnarray}
We have introduced the notation $B$ for the two loop coefficient which has not
yet been computed with the effective theory. From the results in Ref.~\cite{Ira}
we are led to expect that $B= C_A (67/18-\pi^2/6)-5 n_f/9$.

Since we wish to run down from $\mu=m_b$ and $\bn\cdot P\sim m_b$ it is
convenient to introduce the scale $m_b$ into the anomalous dimensions.  Writing
$\ln(\mu/\bn\cdot p)=\ln(\mu/m_b)-\ln(\bnP)$ and noting that the second
logarithm is not large, Eq.~(\ref{adim1}) can be written as
\begin{eqnarray} \label{adim2}
 \gamma_{LO} &=& -\frac{\alpha_s(\mu) C_F}{\pi} 
  \ln\bigg( \frac{\mu}{m_b} \bigg) \,, \nn\\
 \gamma_{NLO} &=& -\frac{\alpha_s(\mu) C_F}{2 \pi} \left[
   \frac{5}{2}- 2 \ln(\bnP)  \right] -2 C_F 
   B\,\frac{\alpha_s^2(\mu)}{(2\pi)^2}\ln\Big( \frac{\mu}{m_b}\Big) \,. 
\end{eqnarray}
Using the one-loop running for $\alpha_s(\mu)$ the LO solution of the RGE is
\begin{eqnarray} \label{LOC}
 \ln\bigg[\frac{C_i(\mu)}{C_i(m_b)}\bigg] =
 \frac{f_0(z)}{\alpha_s(m_b)} = -\frac{4\pi C_F}{\beta_0^2\:\alpha_s(m_b)}
 \:\Big[ \frac{1}{z} -1 + \ln z \Big] \,,
\end{eqnarray}
where $\beta_0 = 11/3 C_A-2/3 n_f$ and
\begin{eqnarray} \label{z}
  z = {\alpha_s(\mu)\over \alpha_s(m_b)} = \frac{2\pi}{2\pi+ {\beta_0}
  \alpha_s(m_b)\ln(\mu/m_b) }\,. 
\end{eqnarray}
Eq.~(\ref{LOC}) sums the LL series of Sudakov logarithms between $Q$ and
$Q\lambda$.  At NLL order we include the $\gamma_{NLO}$ term in the anomalous
dimension and the two-loop running of $\alpha_s(\mu)$ in $\gamma_{LO}$ and find
the following correction to Eq.~(\ref{LOC}) :
\begin{eqnarray} \label{NLOC}
 \ln\bigg[\frac{C_i(\mu)}{C_i(m_b)}\bigg]\bigg|_{NLO} = f_1(z,\bnP) 
 &=& -\frac{ C_F\beta_1}{\beta_0^3} \Big[ 1 -z + z\ln z -\frac12 \ln^2 z
  \Big] \\ 
 && + \frac{C_F}{\beta_0} \Big[ \frac52 - 2 \ln(\bnP) \Big] \ln z
  - \frac{2 C_F B}{\beta_0^2}  \Big[ z -1- \ln z \Big] \,. \nn
\end{eqnarray}
Here $\beta_1 = 34 C_A^2/3-10 C_A n_f/3-2 C_F n_f$, and $z$ is still given by
Eq.~(\ref{z}).  It is easy to see that the result in
Eq.~(\ref{NLOC}) is suppressed by an extra $\alpha_s(m)$ relative to the result
in Eq.~(\ref{LOC}).  Also it is clear that to systematically sum the
next-to-leading log series the two loop coefficient $B$ is required.

Combining Eqs.(\ref{LOC}) and (\ref{NLOC}) the final result at a scale $\mu\sim
Q\lambda$ is
\begin{eqnarray} \label{sln1}
   C_i(\mu) &=& C_i(m_b) \exp\Bigg[\frac{f_0(z)}{\alpha_s(m_b)} 
  + f_1(z) \bigg]  \,.
\end{eqnarray}
For $i=\{1,2,3,4,6,7,9,10,11\}$ the matching starts at tree level and from
Table~\ref{table_logs} we see that for the LL and NLL solutions the value
$C_i(m_b)=1$ should be used in Eq.~(\ref{sln1}).  The coefficients
$C_{5,8,12}(m_b)$ are zero at tree level, and inserting their one-loop
matching values from Eq.~(\ref{matchC}) into Eq.~(\ref{sln1}) gives their LL and
NLL series of logarithms.

\section{Applications }

\subsection{Inclusive decays} \label{incl}

It is well known that the OPE for heavy-to-light decays converges only for
sufficiently inclusive variables. If the available phase space is restricted
such that only a few resonances contribute to the decay the assumption of local
duality no longer holds and the OPE fails. If, however, phase space is
restricted such that highly energetic jets with small invariant mass dominate
the decay, only a subset of terms in the OPE are enhanced. It is possible to
resum this subset of terms into a universal structure
function~\cite{structfun}. In the same region of phase space large Sudakov
logarithms spoil the perturbative expansion and thus have to be summed as
well. This summation was carried out for the endpoint of the leptonic energy
spectrum in inclusive $B\to X_u\ell\bar\nu$ decays and the endpoint of the
photonic energy spectrum in inclusive $B\to X_s\gamma$ decays~\cite{Ira,Ira2}
using perturbative factorization~\cite{path}. In a subsequent work~\cite{LLR}
endpoint logarithms in the hadronic mass spectrum of $B\to X_u\ell\bar\nu$
decays were summed within the same approach. In \cite{BFL} it was shown that the
result for $B \to X_s \gamma$ can be reproduced using the effective field
theory.

In this section we consider $B\to X_s\gamma$ and $B\to
X_u\ell\bar\nu$ decays at the endpoint of the photonic energy and
hadronic invariant mass spectrum, respectively. We define kinematic
variables 
\begin{equation}
  s_0 = \frac{p_u^2}{m_b^2}\,, \qquad h = \frac{2 v\cdot p_u}{m_b} \,, 
\end{equation}
for $B \to X_u \ell\bar\nu$ decays where $p_u$ is the momentum of the $u$ quark.
For $B \to X_s \gamma$ decays we define
\begin{eqnarray}
 x = \frac{2 v\cdot q}{m_b} \,,
\end{eqnarray}
where $q$ is the photon momentum. The endpoint regions mentioned above
correspond to $1-x \sim s_0/h \sim \Lambda_{QCD}/m_b$. Thus, the invariant mass
of the light jet is of order $\sqrt{m_b\Lambda_{QCD}}$, and the power counting
parameter satisfies $\lambda^2 \sim \Lambda_{QCD}/m_b$. At tree level,
integrating out collinear modes by performing an OPE in the collinear-soft
theory and matching onto soft operators gives
\begin{eqnarray} \label{drmoms}
 \frac{d\widehat{\Gamma}_s}{dx} &\equiv& \frac{1}{\Gamma^{(0)}_s}
 \frac{d\Gamma_s}{dx} = \langle B |\, O(x)\,  |B \rangle \nn \\[5pt]
 \frac{d^2\widehat{\Gamma}_u}{dz dh} &\equiv&
   \frac{1}{\Gamma^{(0)}_u}\frac{d^2\Gamma_u}{dz dh} 
   =  2h^2 (3-2h) \langle B |\, O(z)\, | B \rangle \,,
\end{eqnarray}
where $z =1-s_0/h$. Here we defined the tree level decay rates in
the parton model
\begin{eqnarray}
 \Gamma^{(0)}_u = {G^2_F \over 192 \pi^3} |V_{ub}|^2 m^5_b\,,
 \qquad
 \Gamma^{(0)}_s = {G^2_F \over 32 \pi^4} |V^*_{ts}V_{tb}|^2\,\alpha_{\rm em}\, 
 [C^{\rm full}_7]^2
  m^5_b \,,
\end{eqnarray}
where $C_7^{\rm full}$ is the Wilson coefficient of the weak operator mediating
the $b \to s$ radiative transition\cite{GSWS} and we neglect contributions from
operators other than $O_7^{\rm full}$.  The operator appearing in
Eq.~(\ref{drmoms}) is defined as~\cite{structfun}
\begin{eqnarray}
  O(y) = \bar h_v \,\delta(i\hat{D}_+ + 1-y) \,h_v\,,
\end{eqnarray}
where the covariant derivative $\hat{D}_+= D_+/m_b$ includes only soft
gluons. The matrix element of this operator between $B$-meson states is the
light-cone structure function of the $B$ meson.  At higher orders in
perturbation theory the differential decay rates can be expressed as
convolutions of short distance coefficients with the operator $O(y)$. Defining
moments of the decay rates
\begin{eqnarray} \label{Glow}
 {d \widehat{\Gamma}_u (N) \over d h} &=& \frac{1}{\Gamma_u^{(0)}} \int_0^1 
  \! dz \,z^{N-1} \frac{d^2\Gamma_u}{dz dh} \,,\nn\\[5pt]
 \widehat{\Gamma}_s (N) &=&\frac{1}{\Gamma_s^{(0)}}  \int_0^1 \! dx\, x^{N-1} 
  \frac{d\Gamma_s}{dx} \,,
\end{eqnarray}
undoes the convolution, and makes comparison to existing results in the
literature straightforward.  Let  $\mu_0 \sim Q\lambda$ be the scale where 
collinear modes are integrated out. At this scale the moments of the decay 
rates are
\begin{eqnarray} \label{sumedmom}
 {d \widehat{\Gamma}_u (N) \over d h} &=&  2 h^2 (3-2 h) C(\mu_0,h m_b)^2
 \langle\, O(N;\mu_0)\, \rangle \,, \nn \\[5pt]
\widehat{\Gamma}_s (N) &=&
 C(\mu_0,m_b)^2\ \langle\, O(N;\mu_0)\, \rangle\,,
\end{eqnarray}
where the operator $O(N;\mu_0)$ is defined as
\begin{eqnarray}
 O(N;\mu_0) = \int_0^1 \! dy \,y^{N-1} O(y;\mu_0) \,.
\end{eqnarray}
Various coefficients $C_i(\mu,\bn\cdot p)$ can contribute to the decay rates in
Eq.~(\ref{sumedmom}).  However, at NLL order we only need the tree level
matching at $\mu=m_b$ in Eq.~(\ref{Ctree}). Furthermore, at the scale
$\mu_0=m_b/\sqrt{N}$ large logarithms are not introduced when matching onto the
operator $O(N,\mu_0)$~\cite{BFL}. At this scale to NLL order we therefore only
need tree level matching onto the operator $O(N,m_b/\sqrt{N})$.  Since between
$\mu =m_b$ and $\mu=m_b/\sqrt{N}$ all coefficients $C_i(\mu,\bn\cdot p)$ have
universal running the result can be written in terms of a single coefficient
$C(\mu,\bn\cdot p)$, where $C(m_b,\bn\cdot p)=1$ and runs according to
Eq.~(\ref{sln1}).

The running of $C(\mu,\bn\cdot p)$ does not reproduce the full set of Sudakov
logarithms because at $\mu=m_b/\sqrt{N}$ there are additional large logarithms
in the matrix element of $O(N,\mu)$. It has been shown that these additional
logarithms arise from purely soft gluons and can be summed by running from
$\mu=m_b/\sqrt{N}$ to $\mu=m_b/N$~\cite{BFL}. However, taking the ratio of the
decay rates in Eq.~(\ref{Glow}) these matrix elements cancel:
\begin{equation}
 \frac{1}{\widehat{\Gamma}_s (N)}\, \frac{d \widehat{\Gamma}_u (N)}{ d h} = 
 2h^2 (3-2h) \left[ \frac{C(\mu_0,m_b h)}{C(\mu_0,m_b)} \right]^2\,. 
\end{equation}
Thus, all the Sudakov logarithms in the ratio of rates are calculable from the
running of the Wilson coefficients in the collinear soft theory.  Using
Eq.~(\ref{sln1}) this leads to
\begin{equation}
 \frac{1}{\widehat{\Gamma}_s (N)}\, \frac{d \widehat{\Gamma}_u (N)}{ d h} = 
 2 h^2 (3-2h) \:
 {\rm exp}\left( -{4 C_F \over \beta_0} \ln(h) \ln(z) \right) \,, 
\end{equation}
where $z(\mu) = \alpha_s(\mu)/\alpha_s(m_b)$ is evaluated at
$\mu=m_b/\sqrt{N}$. This result agrees with Ref.~\cite{LLR}.

\subsection{Exclusive decays}  \label{excl}

As another application of the results obtained in sections II and III, we
investigate exclusive heavy to light decays. The nonperturbative physics of such
decays is given in terms of form factors. For $B$ decays to pseudoscalar and
vector mesons they are conventionally defined as
\begin{eqnarray} \label{fullff}
\langle P(p)|\bar q \, \gamma^\mu b |\bar{B}(p_b)\rangle &=&
f_+(q^2)\left[p_b^\mu+p^{\mu}-\frac{m_B^2-m_P^2}{q^2}\,q^\mu\right]
+f_0(q^2)\,\frac{m_B^2-m_P^2}{q^2}\,q^\mu, \nn \\
\langle P(p)|\bar q \, i\sigma^{\mu\nu} q_\nu b|\bar{B}(p_b) \rangle &=&
-\frac{f_T(q^2)}{m_B+m_P}\left[q^2(p_b^\mu+p^{\mu})-
(m_B^2-m_P^2)\,q^\mu\right], \nn \\
\langle V(p,\epsilon^\ast)| \bar q \gamma^\mu b | \bar{B}(p_b) \rangle &=&
 \frac{2V(q^2)}{m_B+m_V} \,i\epsilon^{\mu\nu\rho\sigma}
 \epsilon^{\ast}_\nu \, (p_b)_\rho\: p_\sigma, \nn \\
\langle V(p,\epsilon^\ast)| \bar q \gamma^\mu\gamma_5 b | \bar{B}(p_b) 
\rangle &=&
  2m_VA_0(q^2)\,\frac{\epsilon^\ast\cdot q}{q^2}\,q^\mu + 
  (m_B+m_V)\,A_1(q^2)\left[\epsilon^{\ast\mu}-
  \frac{\epsilon^\ast\cdot q}{q^2}\,q^\mu\right] \nn \\
  &&\hspace*{-2cm}
-\,A_2(q^2)\,\frac{\epsilon^\ast\cdot q}{m_B+m_V}
 \left[p_b^\mu+p^{\mu} -\frac{m_B^2-m_V^2}{q^2}\,q^\mu\right], \nn\\
\langle V(p,\epsilon^\ast)| \bar q i\sigma^{\mu\nu}q_\nu b | \bar{B}(p_b)
\rangle &=&
  -2\,T_1(q^2)\,i\epsilon^{\mu\nu\rho\sigma}\epsilon^{\ast}_\nu\, 
  (p_b)_\rho\: p_\sigma, \nn\\
\langle V(p,\epsilon^\ast)| \bar q i\sigma^{\mu\nu} \gamma_5 q_\nu b | 
 \bar{B}(p_b) \rangle &=&
 T_2(q^2)\left[(m_B^2-m_V^2)\,\epsilon^{\ast\mu}-(\epsilon^\ast\cdot
 q)\,(p_b^\mu+p^{\mu})\right]
     \nn \\
 && \hspace*{-2cm} +\,T_3(q^2)\,(\epsilon^\ast\cdot
 q)\left[q^\mu-\frac{q^2}{m_B^2-m_V^2}(p_b^\mu+p^{\mu})\right], 
\end{eqnarray}
where $q=p_b-p$. For decays in which the final light meson has large energy we
can use the effective theory to gain additional information on these
form factors.  Using Eq.~(\ref{ctrnsfm}) the matrix elements in the full theory
are given by matrix elements in the effective theory
\begin{eqnarray} \label{parts}
 \langle M | \bar{q} \,\Gamma b | B \rangle \to \sum_i C_i(\mu) \langle 
 M_{n,P} | \bar \chi_{n,P} \Gamma_i  h_v | H_v \rangle\Big|_\mu + \Delta 
  F_\Gamma\,.
\end{eqnarray}
Here $M = P,V$ correspond to the light pseudoscalar and vector meson states in
the full theory, and $M_{n,P}$ and $H_v$ are the states of the light and the
heavy mesons in the effective theory, respectively.  The first term in
Eq.~(\ref{parts}) is the soft contribution, while the second term indicates the
so-called hard contributions~\cite{Yao,beneke}.  For the soft form factor the
offshellness of the light quark $p_q^2=2 E k_+$ where $k_+\sim \Lambda_{\rm
QCD}$, thus $\lambda^2 \sim \Lambda_{\rm QCD}/m_b$, just as for the inclusive
decays. The $\Delta F_\Gamma$ term in Eq.~(\ref{parts}) involves interactions
where a collinear gluon is exchanged with the spectator in the $B$ meson. In
Ref.~\cite{beneke} it was argued that these spectator effects are the same order
in $\lambda$ and $1/m_b$ as the soft contributions, but can be regarded as being
suppressed by a power of $\alpha_s(\sqrt{m_b \Lambda_{\rm QCD}})$.  They are
therefore just as or more important than the one-loop corrections to the
matching coefficients $C_i(\mu)$ given in Eq.~(\ref{matchC}).  Here we will
apply the effective theory to the soft contributions and leave the hard
spectator contributions for future investigation.

In Ref.~\cite{french}, Charles et al. showed that in heavy to light decays, in
which both the heavy and light quark interact solely via soft gluons, there are
only three independent matrix elements.  Charles et al. derived their result by
combining HQET with LEET and using the fact that the HQET spinors describing
heavy quarks, $h_v$ and the LEET spinors describing highly energetic quarks
interacting with soft gluons, $\xi_{n}$, have only two independent
components. Using the relations $\Slash{v} h_v = h_v$ and $\nslash \xi_{n} = 0$
they showed that at leading order in $1/E$ (where $E$ is the energy of the light
meson) matrix elements of all hadronic currents in LEET are determined by only
three independent functions.  Unfortunately, LEET is not sufficient to describe
heavy to light decays because it omits interactions with collinear gluons.

However, as pointed out in section II the spinors in the effective theory
describing highly energetic quarks interacting with {\it both} soft and
collinear gluons still have two components.  In Eq.~(\ref{ctrnsfm}) we see that
there are only four independent heavy to light currents in the collinear soft
effective theory.  For decays to pseudoscalar mesons like the $\pi$ and $K$ the
matrix elements of these currents are
\begin{eqnarray} \label{P1}
\langle P_n | \bar\chi_{n,P}\, h_v | H_v \rangle &=& 2 E \zeta(E)\,,\nn\\
\langle P_n | \bar\chi_{n,P}\, \gamma^5\, h_v | H_v \rangle &=& 0 \,, \nn\\
\langle P_n | \bar\chi_{n,P}\, \gamma^\mu_\perp\, h_v | H_v \rangle &=& 0 \,,
\end{eqnarray}
while for decays to vector mesons such as $\rho$ and $K^*$ they are
\begin{eqnarray} \label{V1}
\langle V_n | \bar\chi_{n,P}\, h_v | H_v \rangle &=& 0\,, \nn\\
\langle V_n | \bar\chi_{n,P}\, \gamma^5\, h_v | H_v \rangle &=& 
  -2 m_V\, \zeta_{||}(E)\: v\cdot \epsilon^* \,,\nn\\ 
\langle V_n | \bar\chi_{n,P}\, \gamma^\mu_\perp\, h_v | H_v \rangle &=& 
  2 E \zeta_\perp(E)\,  i\,\epsilon^{\mu\nu}_\perp \epsilon^*_\nu \,,
\end{eqnarray}
where $\epsilon^{\mu\nu}_\perp=\epsilon^{\mu\nu\sigma\tau} v_\sigma n_\tau$ and
we are using relativistic normalization for all effective theory states.
Thus, there are still only three linearly independent soft form factors in the
complete effective theory.  Together with Eq.~(\ref{ctrnsfm}) these matrix
elements determine that at tree level the heavy to light form factors are
\begin{eqnarray} \label{ffrel}
\begin{array}{llllll}
& f_+(q^2)=\zeta(E)\,, && f_0(q^2)=2 \hat{E} \zeta(E)\,,&& f_T(q^2)=\zeta(E)\,,\\
& A_1(q^2)=2\hat{E}\,\zeta_\perp(E)\,,\qquad && A_2(q^2)=\zeta_\perp(E)\,, 
 && V(q^2)=\zeta_\perp(E)\,,\label{V} \\
& T_1(q^2)=\zeta_\perp(E)\,,&& T_2(q^2)=2 \hat{E}\,\zeta_\perp(E)\,,\qquad
 && T_3(q^2)=\zeta_\perp(E)\,, \\
& A_0(q^2)= \zeta_{||}(E) \,.
\end{array}
\end{eqnarray}
In deriving these relations we have dropped terms suppressed by $m_{P,V}/E$
since these corrections are expected to be just as large as $\lambda$ suppressed
power corrections which are not included.  Thus, $\zeta_{||}(E)$ only appears in
the purely longitudinal form factor $A_0(q^2)$.  Taking this into account our
results are in agreement with Ref.~\cite{french}.

From the results in section~\ref{match} we can obtain some more information on
the heavy to light form factors. The results of Eqs.~(\ref{ctrnsfm}) and
(\ref{matchC}) determine the perturbative corrections to Eq.~(\ref{ffrel}).
Hard corrections do not break the symmetry relations between effective theory
matrix elements, but do change the relation between form factors in the full and
effective theory.  We find
\begin{eqnarray} \label{ffC}
 f_+(q^2) &=& \zeta(E) \left[ C_4 + \hat{E} C_5 \right] \,, \nn\\ 
 f_0(q^2) &=& \zeta(E) 2 \hat{E}\left[C_4 + C_5 (1-\hat{E}) \right] \,,  \nn\\ 
 f_T(q^2) &=& \zeta(E) C_{11} \,,  \nn\\ 
 A_1(q^2) &=& \zeta_\perp(E) 2 \hat{E} C_3  \,, \nn \\ 
 A_2(q^2) &=& \zeta_\perp(E) C_3 \,, \\ 
 V(q^2) &=& \zeta_\perp(E) C_3  \,,  \nn\\ 
 T_1(q^2) &=& \zeta_\perp(E) C_9  \,,  \nn\\ 
 T_2(q^2) &=& \zeta_\perp(E) 2 \hat{E} C_9  \,,  \nn\\ 
 T_3(q^2) &=& \zeta_\perp(E)  C_9 \nn \,,\\
 A_0(q^2) &=& \zeta_{||}(E) \left[ C_4 + C_5(1-\hat{E}) \right] \nn \,.
\end{eqnarray}
where $C_i=C_i(2\hat E)$ and we have used the helicity relations given below
Eq.~(\ref{matchC}).  In Ref.~\cite{hiller} it was pointed out that the ratios,
$V/A_1$ and $T_1/T_2$ do not receive perturbative corrections due to the fact
that interactions which flip the helicity of the energetic quark are suppressed
by $1/E$.  From Eq.~(\ref{ffC}) we see that in fact at leading order in
$\lambda$ the soft contributions to the form factors $\{A_1,A_2,V\}$ and
$\{T_1,T_2,T_3\}$ are related to all orders in $\alpha_s$.  Furthermore, since
the RGE's for all currents are identical any ratio of soft form factors are
independent of Sudakov logarithms.

At one loop the hard corrections to ratios of the form factors in
Eq.~(\ref{fullff}) were previously calculated in Ref.~\cite{beneke}.  Since the
authors used LEET as their effective theory their matching calculation was not
infrared safe and the overall normalization of the low energy matrix elements
was unknown. However, it was noted that this problem cancels out of ratios of
form factors because the infrared divergences in the full theory are
universal. Our results in Eq.~(\ref{ffC}) do not suffer from this problem
because the collinear-soft effective theory has the same infrared divergences as
QCD. Taking ratios of the form factors in Eq.~(\ref{ffC}), substituting the
results in Eq.~(\ref{matchC}), and expanding in $\alpha_s(m_b)$, our results for
the hard corrections to the soft form factors agree with those of
Ref.~\cite{beneke}.

As an application, consider the zero in the forward-backward asymmetry of the
rare decay $B \to K^* \ell^+ \ell^-$, which gives a relation between the Wilson
coefficients $C_9^{\rm full}$ and $C_7^{\rm full}$~\cite{morozumi,burdman}
\begin{eqnarray}
{\rm Re}\Bigg[\frac{ C_9^{\rm full}(s_0) }{C_7^{\rm full} }\Bigg] = -\frac{m_b}{s_0} 
\left[\frac{T_2(s_0)}{A_1(s_0)}\left(m_B-m_{K^*}\right) +
\frac{T_1(s_0)}{V(s_0)}\left(m_B+m_{K^*}\right)\right] \,, 
\end{eqnarray}
where here $s_0\sim 3\,{\rm GeV}$ is the value of $q^2$ where the asymmetry
vanishes. It was noted in Ref.~\cite{burdman} that in the ratio of soft form
factors the effective theory form factors cancel. Ignoring again the hard
spectator contributions and the higher order effect of the mass of the $K^*$ we
find
\begin{eqnarray} \label{c9c7}
{\rm Re}\Bigg[\frac{ C_9^{\rm full}(s_0) }{C_7^{\rm full} }\Bigg] &=& 
 - m_B\, \frac{m_b}{s_0}    \left[\frac{2C_9(m_b)}{C_3(m_b)} 
 \right] \nn\\ &=& - 2 m_B\, \frac{m_b}{s_0}  \left( 1+\frac{\alpha_s(m_b)
C_F}{4\pi} \ln(2\hat{E}) \frac{2\hat{E}}{1-2\hat{E}}\right)\,,
\end{eqnarray}
where the perturbative correction from the soft form factor is in agreement with
Ref.~\cite{beneke}. There are additional order $\alpha_s$ corrections to
Eq.~(\ref{c9c7}) from collinear gluon exchange with the spectator in the B,
which can be found in Ref.~\cite{beneke}.  Although Sudakov logarithms do not
effect the ratio of purely soft form factors, they may suppress the soft
contribution relative to that from collinear gluon exchange.  

\section{Conclusions}

In this paper we investigated in detail the collinear-soft effective theory,
which describes highly energetic particles with low invariant mass. The degrees
of freedom in this theory consist of collinear quarks and gluons with momenta
scaling as $k_c = Q(\lambda^2, 1,\lambda)$ and soft gluons with momenta scaling
as $k_s = Q(\lambda^2,\lambda^2,\lambda^2)$. We gave a detailed derivation of
the collinear-soft Lagrangian with the intent of making it straightforward to go
to subleading orders in $\lambda$. In addition we derived the effective theory
heavy-to-light current at order $\lambda^0$. For decays of heavy particles there
are regions of phase space where this theory applies, namely when the hadronic
decay products are light and are produced with large energy. The currents
mediating these decays are given by four linearly independent operators in the
effective theory. We performed the matching onto these operators at the one loop
level and calculated their renormalization group evolution from the hard scale
$Q\sim m_b$ to the intermediate scale $Q\lambda$.

We considered two applications of the collinear-soft theory: inclusive and
exclusive decays. In the inclusive case we focused our attention on the
radiative decay $B \to X_s \gamma$ and the semileptonic decay $B \to X_u \ell
\bar\nu$ in the endpoint region of large photon energy and of low hadronic
invariant mass, respectively.  At leading order the OPE in the effective theory
gives a bilocal operator whose matrix element is the universal nonperturbative
light cone structure function of the $B$ meson. As is well known, in the ratio
of large moments of these two decays this structure function cancels. As a
consequence the Sudakov logarithms in this ratio are entirely determined by the
running in the collinear-soft theory as discussed in Section~\ref{incl}. Our
result is in agreement with previous literature~\cite{Ira,LLR}.

For exclusive decays we investigated the relationship amongst form factors in
the large energy limit of QCD. In Ref.~\cite{french} it was shown using LEET
that there are only three independent soft form factors at leading order in an
expansion in inverse powers of the energy of the light quark. However, since
LEET does not include collinear gluons it does not correctly reproduce the IR
logarithms of QCD, and the relevance of this result is not immediately
obvious. We showed that the presence of collinear gluons does not spoil the
relations among the soft form factors, therefore establishing these results in
the large energy limit of QCD. Finally we used the one loop matching of the
currents in the effective theory to relate the full theory form factors to the
three independent matrix elements in the effective theory. Our analysis confirms
the corresponding results in Ref.~\cite{beneke}, but with an infrared safe
definition of the matching coefficients.

\acknowledgements

We would like to thank Michael Luke for collaboration on early stages of this
project and Ira Rothstein for numerous discussions. We would also like thank
Mark Wise for discussion on the helicity invariance.  C.B., S.F., and I.S. would
like to thank the INT at the University of Washington for their hospitality at
an early stage of this work. S.F. would like to thank the Caltech theory group
for their hospitality while part of this work was completed. D.P. is grateful to
the National Center of Theoretical Sciences, R.O.C.\ for hospitality during the
final phase of this work.  This work was supported in part by the Department of
Energy under grants DOE-FG03-97ER40546 and DOE-ER-40682-143 and by NSERC of
Canada.


\appendix
 
\section{Collinear Gauge Transformations} \label{app_gauge}

In this appendix we discuss the collinear gauge invariance in the soft-collinear
effective theory. For simplicity we will restrict ourselves to the abelian
case. From the general set of gauge transformations $U(x) = e^{i\alpha(x)}$,
where
\begin{eqnarray} \label{trn1}
  \psi(x) \to U(x)\psi(x)\,, \qquad\quad
  A_\mu(x) \to A_\mu(x) - \frac{i}{g}U^\dagger (x) \partial_\mu U(x) \,,
\end{eqnarray}
the collinear transformations belong to a subset where $\partial_\mu \alpha(x)$
scales like a collinear momentum. To make this scaling explicit we decompose
an arbitrary collinear gauge transformation as
\begin{eqnarray} \label{cgauge}
  U(x) \equiv \int d^4Q\: e^{iQ\cdot x}\: \beta(Q) 
  = \sum_{\tilde Q}\: e^{i\tilde Q\cdot x}\: \beta_Q(x^-) \,,
\end{eqnarray}
where $(\bn\cdot Q,Q_\perp,n\cdot Q)\sim (\lambda^0,\lambda,\lambda^2)$ and the 
sum is over $\tilde Q=(\bn\cdot Q,Q_\perp)$. For notational convenience we will
suppress the dependence of $\beta_Q$ on $x^-$ henceforth.

In Section~II the full quark field was decomposed into components $\xi_{n,p}(x)$
which no longer depend on the large phases $e^{-i\tilde p\cdot x}$.  Under the
collinear gauge transformation in Eq.~(\ref{cgauge}) we have
\begin{eqnarray} \label{sfermi}
  \sum_{\tilde p} e^{-i\tilde p\cdot x}\xi_{n,p}(x) \to 
 \sum_{\tilde p} \sum_{\tilde Q} e^{-i(\tilde p-\tilde Q)\cdot x}\beta_Q \:
 \xi_{n,p}(x) 
 = \sum_{\tilde p} e^{-i\tilde p \cdot x}\sum_{\tilde Q}  \beta_Q \:
 \xi_{n,p+Q}(x) \,.
\end{eqnarray}
Up to terms suppressed by powers of $\lambda$ the $x$ dependence of $\xi_{n,p}$
can be ignored and the Fourier components must agree, so
\begin{eqnarray} \label{fermi}
  \xi_{n,p}(x) \to \sum_{\tilde Q} \beta_{Q-p}\: \xi_{n,Q}(x)\,.
\end{eqnarray}
Thus, the collinear gauge invariance simply corresponds to a
``reparameterization'' invariance of the theory under changes to the effective
theory labels. Similarly, for the collinear gluon field with label $\tilde q$ we
find
\begin{eqnarray} \label{sgauge}
  \sum_{\tilde q} e^{-i\tilde q\cdot x} A^\mu_{n,q}(x) \to 
 \sum_{\tilde q}  e^{-i\tilde q\cdot x} A^\mu_{n,q}(x) + 
 \frac{1}{g} \sum_{\tilde R} e^{-i\tilde R\cdot x} \sum_{\tilde Q} \beta^*_{R+Q} 
 \Big[ \beta_Q Q^\mu -i\partial^\mu \beta_Q \Big]  \,,
\end{eqnarray}
so the components transform as
\begin{eqnarray}\label{gauge}
 A_{q}^\mu \to A_{q}^\mu + \frac{1}{g}\sum_{\tilde Q} \beta_{Q+q}^*\, 
  \Big[ Q^\mu \beta_Q -i\partial^\mu \beta_Q \Big] \,.
\end{eqnarray}

Using the transformation properties (\ref{fermi}) and (\ref{gauge}) for the
collinear quark and gluon field respectively, it is easy to see that the
soft-collinear effective Lagrangian in Eq.~(\ref{pre2Lc}) is gauge invariant. To
see this, it is sufficient to note that the following combination of collinear
fields transforms in the same manner as the collinear quark field in
Eq.~(\ref{sfermi}),
\begin{eqnarray}
 \sum_{\tilde p} e^{-i\tilde p\cdot x} \left( \tilde p^\mu  + 
 \frac{\bn^\mu}{2} in\cdot\partial
 + g\sum_{\tilde q} e^{-i\tilde q\cdot x} A^\mu_{n,q} \right) \xi_{n,p}\,.  
\end{eqnarray}
The derivation is somewhat tedious, so we will not display the details. However,
we note that to derive this result it is necessary to make use of the unitarity
of the gauge transformation, $U^\dagger(x) U(x) =1$ which implies
\begin{eqnarray} \label{unitarity}
  \sum_{\tilde P,\tilde Q} \beta_Q\, \beta_P^*\: e^{i(Q-P)\cdot x} = 1\,.  
\end{eqnarray}

Finally, we show that the jet field $\chi_{n,P}$ from Section~III, 
\begin{eqnarray} \label{cJet}
 \chi_{n,P} = \sum_{\tilde p} e^{-i\tilde p\cdot x}\exp\left(
 \sum_{\tilde q} e^{-i\tilde q\cdot x}\: \frac{g\,\bar n\cdot A_{n,q}}
 {\bar n\cdot q}\right) \xi_{n,p}\,,
\end{eqnarray}
is invariant under the collinear gauge transformation in Eq.~(\ref{cgauge}).
We have
\begin{eqnarray} \label{ch1}
 \chi_{n,P} \to \sum_{\tilde p} e^{-ip\cdot x}\exp\left[
 \sum_{\tilde q} e^{-i\tilde q\cdot x} \frac{g\bar n\cdot A_{n,q}}
 {\bar n\cdot q}  + \sum_{\tilde q} \frac{ e^{-i\tilde q\cdot x} }
 {\bar n\cdot q}  \sum_{\tilde Q} \bar n\cdot Q\: \beta_Q\: \beta_{Q+q}^* 
 \right] \sum_R \beta_{R-p}\: \xi_{n,R}\,.
\end{eqnarray}
Comparing Eqs.~(\ref{trn1}) and (\ref{sgauge}) and using
$\bn\cdot\partial\beta_Q =0$ gives
\begin{eqnarray}
 -i\,\bn\cdot\partial\alpha(x) = \sum_{\tilde R,\tilde Q} e^{-i\tilde R\cdot x}
 \beta^*_{R+Q}\, \beta_Q \, (-i\bn\cdot Q ) \,.
\end{eqnarray} 
Integrating this result with respect to $n\cdot x/2$ and taking the exponential
gives the relation
\begin{eqnarray} \label{exprel}
 \exp\Big[ \sum_{\tilde R,\tilde Q} e^{-i\tilde R\cdot x}
 \beta^*_{R+Q}\, \beta_Q \, \frac{\bn\cdot Q }{\bn\cdot R} \Big] 
 = e^{-i\alpha(x)} = \sum_{\tilde Q} e^{-i\tilde Q\cdot x} \beta^*_{Q} \,.
\end{eqnarray}
Substituting Eq.~(\ref{exprel}) into Eq.~(\ref{ch1}) and shifting $\tilde p \to
\tilde R-\tilde p$ leaves
\begin{eqnarray}
 \sum_{\tilde R}  e^{-i\tilde R\cdot x} \exp\left[
 \sum_{\tilde q} e^{-iq\cdot x} \frac{g\bar n\cdot A_{n,q}}{\bar n\cdot q}\right]
 \sum_{\tilde p} e^{i\tilde p\cdot x}
  \beta_p \sum_{\tilde Q} e^{-i\tilde Q\cdot x} \beta^*_Q\: \xi_{n,R} =
 \chi_{n,P} \,,
\end{eqnarray}
where in the last step we have used the unitarity relation in
Eq.~(\ref{unitarity}) and the definition in Eq.~(\ref{cJet}). Thus, the jet field is invariant under a collinear gauge
transformation as expected.

\section{Renormalization of the current in the collinear-soft
theory} \label{app_scren}

In Section III we quoted the renormalization constant of the current
operator $\bar\chi_{n,P} \Gamma h_v$ in the collinear-soft theory
\begin{eqnarray}\label{A1}
Z = 1 + \frac{\alpha_s C_F}{4\pi}\left(
\frac{1}{\epsilon^2} + \frac{2}{\epsilon}\log\frac{\mu}{\bar n\cdot P}
+ \frac{5}{2\epsilon}\right)\,.
\end{eqnarray}
This was obtained by computing the renormalization of the term $\bar\xi_{n,p}
\Gamma h_v$ which is the first term obtained using the expansion of
$\bar\chi_{n,P}$ in Eq.~(\ref{jetq}). However, the renormalization constant
(\ref{A1}) depends only on the large component of the jet momentum $\bar n\cdot
P$, which enters as a label on the jet field $\chi_{n,P}$. This is a nontrivial
consequence of collinear gauge invariance and is essential for a consistent
renormalization of the collinear-soft effective theory.  For example, in the
collinear diagram in Fig.~\ref{fd_eft}c the Wilson coefficient depends on only
the sum of the collinear gluon and quark momentum in the loop.  Thus, it depends
only on $p$ and not on the loop momentum.

In this Appendix we illustrate this property of the current by explicit
calculation of the corresponding renormalization of the one collinear gluon term
in the expansion of $\bar\chi_{n,P} \Gamma h_v$. For simplicity we will work
with an Abelian gauge theory (QED), for which this expansion has been given
explicitly in Eq.~(\ref{J2}) (with $P=p+\sum_i q_i$ for each term)
\begin{eqnarray}\label{A2}
\bar\chi_{n,P} \Gamma h_v = \bar\xi_{n,P}\Gamma h_v -  
\frac{g}{\bar n\cdot q}\bar\xi_{n,p} \bar n\cdot A_{n,q}\Gamma h_v
+ \frac{g^2}{\bar n\cdot q_1 \bar n\cdot q_2}
\bar\xi_{n,p} \bar n\cdot A_{n,q_1}\bar n\cdot A_{n,q_2}\Gamma h_v
+ \cdots\,.
\end{eqnarray}

\begin{figure}[!t]
 \centerline{\mbox{\epsfysize=3.0truecm \hbox{\epsfbox{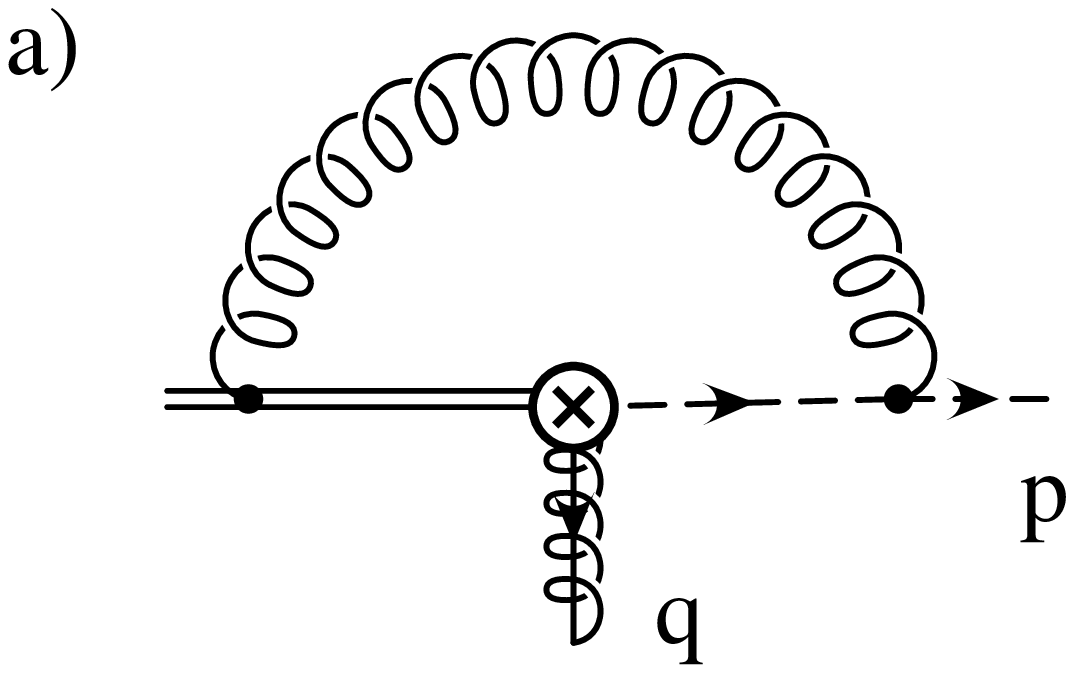}} 
\hspace{2.cm}
  \epsfysize=3.0truecm \hbox{\epsfbox{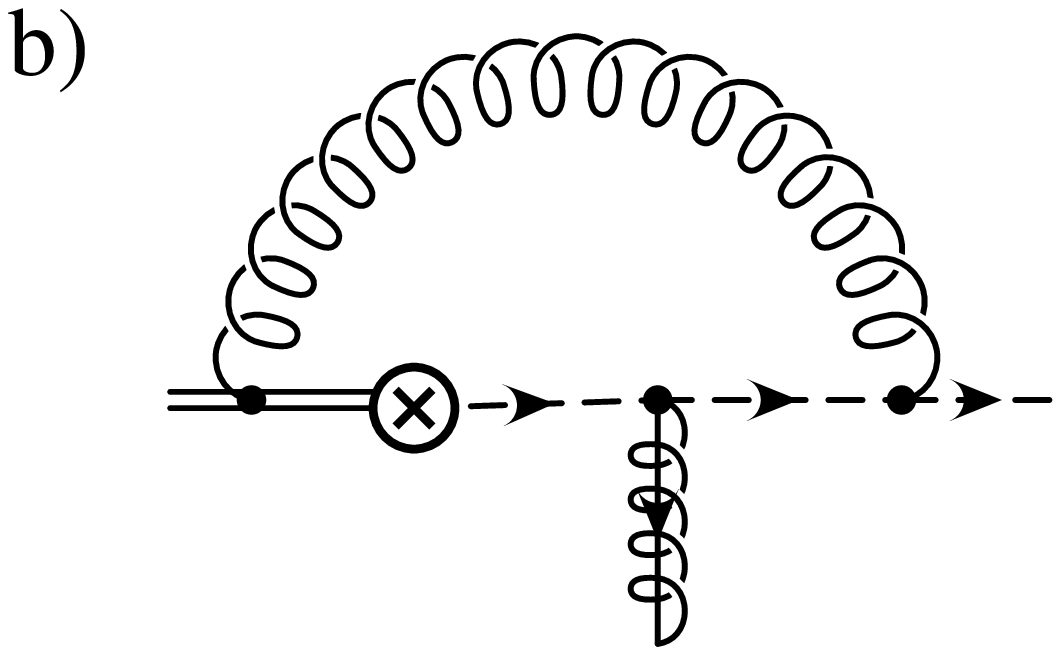}}  }}
 \medskip
{\tighten \caption[1]{One-gluon diagrams contributing to the soft 
gluon renormalization of the current operator 
$\bar \chi_n \Gamma h$ in an abelian gauge theory. The crossed dot
denotes one insertion of the operator $\bar\chi_n\Gamma h_v$.} 
\label{fd10} }  
\end{figure}

\begin{figure}[!t]
 \centerline{\mbox{\epsfysize=2.7truecm \hbox{\epsfbox{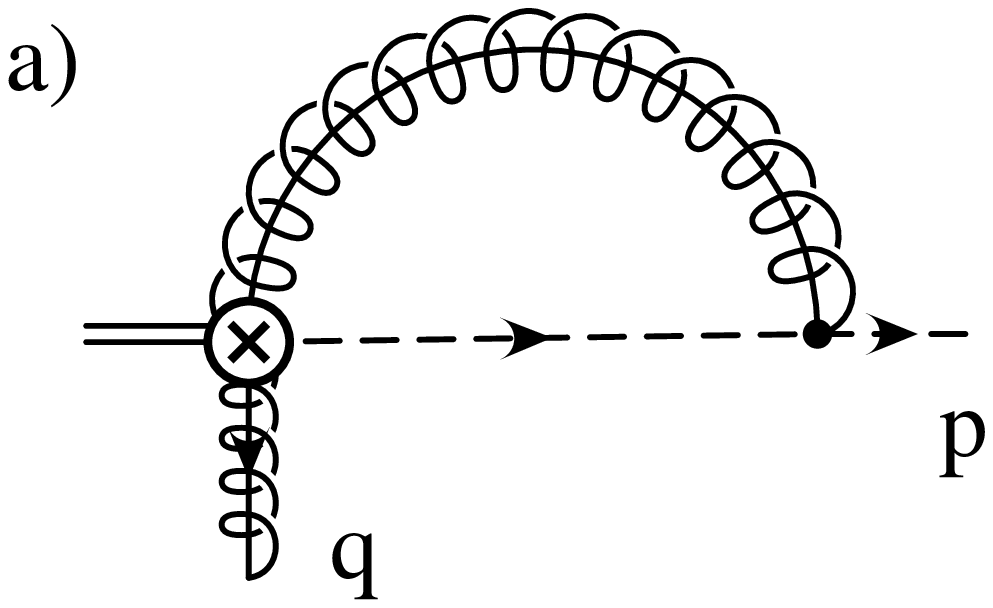}} 
\hspace{1cm}
  \epsfysize=2.7truecm \hbox{\epsfbox{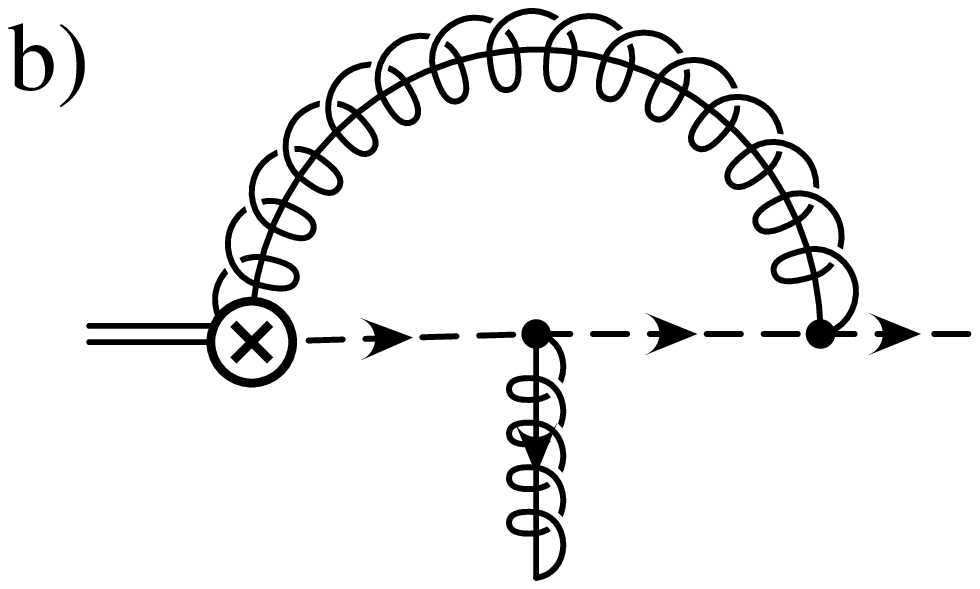}}
\hspace{1cm}
  \epsfysize=2.7truecm \hbox{\epsfbox{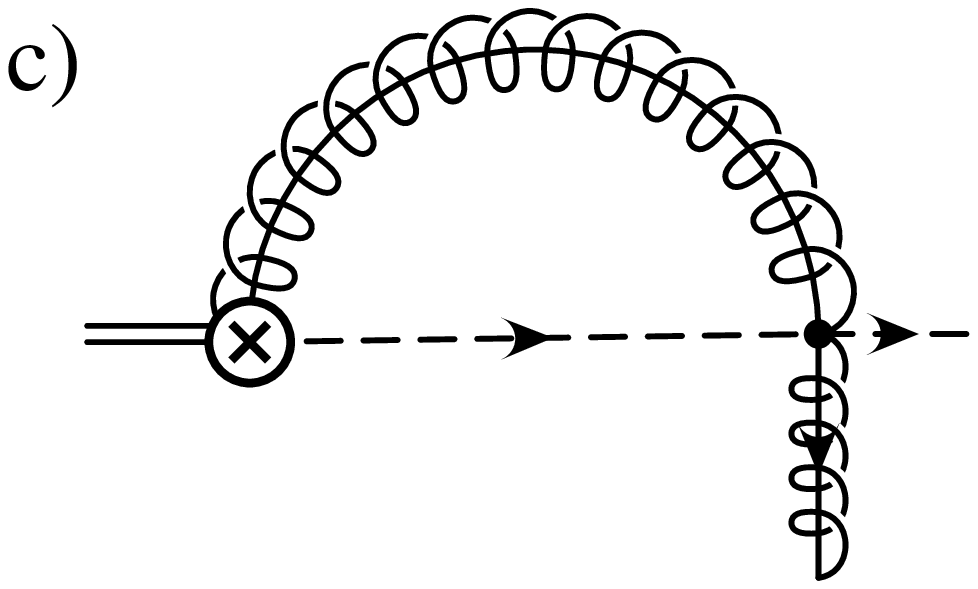}}  }}
 \medskip
{\tighten \caption[1]{Collinear gluon renormalization of the
current operator $\bar \chi_n \Gamma h$ with one external collinear
gluon. The crossed dot denotes one insertion of the operator 
$\bar\chi_n\Gamma h_v$.} \label{fd11} }  
\end{figure}

The diagrams contributing to the renormalization of the second term in this
expansion are shown in Figs.~\ref{fd10} and \ref{fd11}. We will work throughout
in Feynman gauge.  The diagrams in Fig.~\ref{fd10}(a) and Fig.~\ref{fd11}(a)
have been computed already; the external gluon momentum $q$ does not enter the
loop integral, so they can be simply extracted from the corresponding results
for $\bar\xi_{n,P}\Gamma h_v$ (Eqs.~(\ref{efts}), (\ref{eftc})):
\begin{eqnarray}\label{A3}
 \ref{fd10}({\rm a}) + \ref{fd11}({\rm a}) = \langle -
 \frac{g}{\bar n\cdot q}\bar\xi_{n,p} \bar n\cdot A_{n,q}\Gamma h_v
 \rangle \cdot \frac{\alpha}{4\pi}\left(
 \frac{1}{\epsilon^2} + \frac{2}{\epsilon} +
 \frac{2}{\epsilon}\log\frac{\mu}{\bar n\cdot p}
 + \mbox{const}\right)\,.
\end{eqnarray}
Furthermore, upon examining the Feynman rule for the two collinear gluon
coupling in Fig.~\ref{fr1}, one can see that the graph in Fig.~\ref{fd11}(c)
vanishes in Feynman gauge. We will show in the following that the net effect of
the two remaining graphs Fig.~\ref{fd10}(b) and Fig.~\ref{fd11}(b) is to change
$\bar n\cdot p$ in the argument of the logarithm in (\ref{A3}) to $\bar n\cdot
(p+q)$, corresponding to the total momentum $P=p+q$ carried by the jet.

For simplicity we will take the external momenta $p,q$ to be off-shell and to
have vanishing transverse components, $p=(p_+,p_-,0_\perp)$ and
$q=(q_+,q_-,0_\perp)$.  With this choice the soft diagram Fig.~\ref{fd10}(b)
reduces to one term, 
\begin{equation} \label{A4}
 7({\rm b}) = -i g^3  \int\frac{\mbox{d}^d l}{(2\pi)^d}
 \frac{\langle \bar\xi_{n,p} n\cdot A_{n,q}\Gamma h_v
 \rangle\: \mu^{2 \epsilon}}{[n\cdot l-p^2/\bar n\cdot p][n\cdot l-
 (p+q)^2/\bar n\cdot (p+q)] [v\cdot l] \,[ l^2]}\,. 
\end{equation}
The integration is performed most easily in light-cone coordinates
$l=(l^+,l^-,l_\perp)$, where the $l^+$ integral can be done by the method of
residues. We obtain
\begin{equation} \label{A5}
 7({\rm b}) = \langle \bar\xi_{n,p} n\cdot A_{n,q}\Gamma h_v \rangle
 {g^3 \over 8 \pi^2} \frac{1}
 {n\cdot q }\left\{ {1 \over \epsilon}
 \ln\left[ \frac{n\cdot(p+q)}{n\cdot p} \right]
 + {\rm const} \right\} \,.
\end{equation}
Since this graph does not give a contribution to $\langle \bar\xi_{n,p} \bn\cdot
A_{n,q}\Gamma h_v \rangle$, it does not contribute to the renormalization of the
current. However, the resulting divergence seems to require a new truly
non-local operator. We will show that this contribution cancels in the sum of
diagrams.

The collinear graph in Fig.~\ref{fd11}(b) can be written as
\begin{eqnarray}
8({\rm b}) = \langle \bar\xi_{n,p} n\cdot A_{n,q}\Gamma h_v \rangle 
I_1 + \langle \bar\xi_{n,p} \bar n\cdot A_{n,q}\Gamma h_v \rangle I_2 \,,
\end{eqnarray}
where
\begin{eqnarray} \label{A7and8}
I_1 &=& 
2ig^3 \mu^{2\epsilon} \int\frac{\mbox{d}^d l}{(2\pi)^d}
\frac{\bar n\cdot(p+l) \bar n\cdot(p+q+l)}
{\bar n\cdot l(l+p)^2 (l+p+q)^2 l^2}\,,
\\
I_2 &=& 
-2ig^3 \mu^{2\epsilon} \int\frac{\mbox{d}^d l}{(2\pi)^d}
\frac{l_\perp^2}{\bar n\cdot l(l+p)^2 (l+p+q)^2 l^2}\,.
\end{eqnarray}
Once again the integration is simplified by using the method of residues on the
$l^+$ integral.  Explicitly, we find
\begin{eqnarray} \label{A9and10}
I_1 &=& 
-{g^3 \over 8 \pi^2} \frac{1}{n\cdot q}\left\{ {1 \over \epsilon}
\ln\left[ \frac{n\cdot(p+q)}{n\cdot p} \right]
+ {\rm const} \right\} \,,
\\
I_2 &=&
- \frac{g}{\bar n\cdot q}\cdot
\frac{\alpha}{4\pi}\left\{ \frac{2}{\epsilon}
\ln\left[ \frac{\bar n\cdot p}{\bar n\cdot (p+q)}\right] + \mbox{const}
 \right\} \,.
\end{eqnarray}
Thus, the first term in $I_1$ cancels the UV divergence in Eq.~(\ref{A5}), as
required. The divergent term in $I_2$ converts the label in Eq.~(\ref{A3}) from
$\bar n\cdot p$ to $\bar n\cdot (p+q)$. As advertised, the remaining UV
divergence depends only on the total jet momentum $P=p+q$:
\begin{eqnarray}\label{A8}
 \ref{fd10}(a,b) + \ref{fd11}(a,b,c) = \langle -
 \frac{g}{\bar n\cdot q}\bar\xi_{n,p} \bar n\cdot A_{n,q}\Gamma h_v
 \rangle \cdot \frac{\alpha}{4\pi}\left(
 \frac{1}{\epsilon^2} +\frac{2}{\epsilon} + 
 \frac{2}{\epsilon}\log\frac{\mu}{\bar n\cdot P}
 + \mbox{const}\right)\,.
\end{eqnarray}
After adding the contributions from the heavy quark and collinear quark field
wavefunction renormalization, we reproduce the renormalization constant $Z$ in
Eq.~(\ref{A1}) (after taking the color factor $C_F\to 1$).  With similar
techniques we have also checked that this holds for the renormalization of the
term in Eq.~(\ref{A2}) that contains two collinear gluon fields. As argued in
section III, collinear gauge invariance forces all the terms in the sum in
Eq.~(\ref{A2}) to be renormalized in the same way, with a Wilson coefficient
which depends only on the total jet momentum. The explicit calculations in this
Appendix agree with this result.

\newpage

{\tighten

} 


\begin{references}

\bibitem{IW} N. Isgur and M.B. Wise, Phys. Lett. B232 (1989) 113; 
Phys. Lett. B237 (1990) 527.

\bibitem{HQET}
E.~Eichten and B.~Hill,
Phys.\ Lett.\  {\bf B234}, 511 (1990);
H.~Georgi,
Phys.\ Lett.\  {\bf B240}, 447 (1990).

\bibitem{incl}
M.~A.~Shifman and M.~B.~Voloshin,
Sov.\ J.\ Nucl.\ Phys.\  {\bf 41}, 120 (1985);
J.~Chay, H.~Georgi and B.~Grinstein,
Phys.\ Lett.\  {\bf B247}, 399 (1990);
I.~I.~Bigi, M.~Shifman, N.~G.~Uraltsev and A.~Vainshtein,
Phys.\ Rev.\ Lett.\  {\bf 71}, 496 (1993).


\bibitem{dugan} M.~J.~Dugan and B.~Grinstein,
Phys.\ Lett.\  {\bf B255}, 583 (1991).

\bibitem{LEETincons}
U.~Aglietti, G.~Corbo and L.~Trentadue,
Int.\ J.\ Mod.\ Phys.\  {\bf A14}, 1769 (1999);
C.~Balzereit, T.~Mannel and W.~Kilian,
Phys.\ Rev.\  {\bf D58}, 114029 (1998).

\bibitem{BFL}
C.~W.~Bauer, S.~Fleming and M.~Luke,
Phys.\ Rev.\  {\bf D63}, 014006 (2001).


\bibitem{LMR}
M.~E.~Luke, A.~V.~Manohar and I.~Z.~Rothstein,
Phys.\ Rev.\  {\bf D61}, 074025 (2000).

\bibitem{BS} V.~A.~Smirnov, Phys. Lett. {\bf B404}, 101 (1997); M.~Beneke and
V.~A.~Smirnov, Nucl. Phys. {\bf B522}, 321 (1998); V.~A.~Smirnov and
E.~R.~Rakhmetov, Theor. Math. Phys. {\bf 120}, 870 (1999); 
V.~A.~Smirnov,
Phys.\ Lett.\  {\bf B465}, 226 (1999); For applications see:
J.~H.~Kuhn, A.~A.~Penin and V.~A.~Smirnov,
Eur.\ Phys.\ J.\  {\bf C17}, 97 (2000);
M.~Beneke, G.~Buchalla, M.~Neubert and C.~T.~Sachrajda,
Nucl.\ Phys.\  {\bf B591}, 313 (2000).

\bibitem{Ira}  
R.~Akhoury and I.~Z.~Rothstein,
Phys.\ Rev.\  {\bf D54}, 2349 (1996);
G.~P.~Korchemsky and G.~Sterman,
Phys.\ Lett.\  {\bf B340}, 96 (1994).

\bibitem{LLR} A.~K.~Leibovich, I.~Low, and I.~Z.~Rothstein,
Phys. Rev. {\bf D62}, 014010 (2000).

\bibitem{russian}
V.~Chernyak and I.~Zhitnitsky, Nucl.\ Phys.\ {\bf B345}, 137 (1990).


\bibitem{french}
J.~Charles, A.~Le Yaouanc, L.~Oliver, O.~Pene and J.~C.~Raynal,
Phys.\ Rev.\  {\bf D60}, 014001 (1999).

\bibitem{beneke}
M.~Beneke and T.~Feldmann,
Nucl.\ Phys.\  {\bf B592}, 3 (2000).


\bibitem{lightconeQCD}
S.~J.~Brodsky, H.~Pauli and S.~S.~Pinsky,
Phys.\ Rept.\  {\bf 301}, 299 (1998);
R.~Venugopalan,
nucl-th/9808023.
H.~Leutwyler, Nucl.\ Phys.\ {\bf B76}, 413 (1974).

\bibitem{infiniteQCD}
J.~D.~Bjorken, J.~B.~Kogut and D.~E.~Soper,
Phys.\ Rev.\  {\bf D3}, 1382 (1971);
J.~B.~Kogut and D.~E.~Soper,
Phys.\ Rev.\  {\bf D1}, 2901 (1970).

\bibitem{rescale} M.~Luke and A.~V.~Manohar, Phys. Rev. {\bf D55},
4129 (1997); M.~Luke and M.~J.~Savage, Phys. Rev. {\bf D57}, 413
(1998).

\bibitem{lightconeQCD_FR} W.~Zhang and A.~Harindranath, Phys.\ Rev.\ 
{\bf D48}, 4881 (1993).

\bibitem{GSWS}
B.~Grinstein, R.~Springer and M.~B.~Wise,
Phys.\ Lett.\  {\bf B202}, 138 (1988);
B.~Grinstein, M.~J.~Savage and M.~B.~Wise,
Nucl.\ Phys.\  {\bf B319}, 271 (1989).


\bibitem{path}
J.~C.~Collins and D.~E.~Soper,
Nucl.\ Phys.\  {\bf B193}, 381 (1981);
J.~C.~Collins, D.~E.~Soper and G.~Sterman,
Nucl.\ Phys.\  {\bf B261}, 104 (1985); See J.~C.~Collins in
Perturbative Quantum Chromodynamics, Ed. A.~H.~Mueller, 573 (1989) and
references therein. 

\bibitem{structfun}
I.~I.~Bigi, M.~A.~Shifman, N.~G.~Uraltsev and A.~I.~Vainshtein,
Int.\ J.\ Mod.\ Phys.\  {\bf A9}, 2467 (1994);
%
M.~Neubert,
Phys.\ Rev.\  {\bf D49}, 4623 (1994).

\bibitem{Ira2}
A.~K.~Leibovich and I.~Z.~Rothstein,
Phys.\ Rev.\  {\bf D61}, 074006 (2000);
A.~K.~Leibovich, I.~Low and I.~Z.~Rothstein,
Phys.\ Rev.\  {\bf D61}, 053006 (2000).

\bibitem{Yao}
R.~Akhoury, G.~Sterman and Y.~P.~Yao,
Phys.\ Rev.\  {\bf D50}, 358 (1994).

\bibitem{hiller}
G.~Burdman and G.~Hiller,
hep-ph/0011266.

\bibitem{morozumi}
A.~Ali, T.~Mannel and T.~Morozumi,
Phys.\ Lett.\  {\bf B273}, 505 (1991).

\bibitem{burdman}
G.~Burdman,
Phys.\ Rev.\  {\bf D57}, 4254 (1998);
A.~Ali, P.~Ball, L.~T.~Handoko and G.~Hiller,
Phys.\ Rev.\  {\bf D61}, 074024 (2000).



\end{references}
\end{document}